\documentclass[11pt,onecolumn,english,preprintnumbers,amsmath,amssymb,floatfix,nofootinbib]{revtex4}

\usepackage{epsfig}
\usepackage{graphics}
\usepackage{latexsym}
\usepackage{amsmath}
\usepackage{amssymb}
\usepackage{rotating}
\usepackage{subfigure}
\usepackage{bm}
\usepackage{color}
\usepackage{ulem}
\usepackage{booktabs}

\usepackage[colorlinks = true,
linkcolor = magenta,
urlcolor  = blue,
citecolor = red,
anchorcolor = blue]{hyperref}

\begin{document}


\title{ Determination of contributions from residual light charged hadrons to inclusive charged hadrons from $e^+e^-$ annihilation data }


\author{Alireza Mohamaditabar$^{1}$}
\email{A.Mohamaditabar@mail.um.ac.ir}

\author{F. Taghavi-Shahri$^{1}$}
\email{Taghavishahri@um.ac.ir}

\author{Hamzeh Khanpour$^{2,3}$}
\email{Hamzeh.Khanpour@cern.ch}

\author{Maryam Soleymaninia$^{3}$}
\email{Maryam\_Soleymaninia@ipm.ir}

\affiliation {
$^{1}$Department of Physics, Ferdowsi University of Mashhad, P.O.Box 1436, Mashhad, Iran  \\	
$^{2}$Department of Physics, University of Science and Technology of Mazandaran, P.O.Box 48518-78195, Behshahr, Iran    \\
$^{3}$School of Particles and Accelerators, Institute for Research in Fundamental Sciences (IPM), P.O.Box 19395-5531, Tehran, Iran       
 }

\date{\today}

%
\begin{abstract}

In this paper, we present an extraction of the contribution from the ``{\it residual}'' light charged hadrons to the inclusive unidentified light charged hadron fragmentation functions (FFs) at next-to-leading (NLO) and, for the first time, at next-to-next-to-leading order (NNLO) accuracy in perturbative QCD.
Considering the contributions from charged pion, kaon and (anti)proton FFs from recent {\tt NNFF1.0} analysis of charged hadron FFs, we determine the small but efficient {\it residual} charged hadron FFs from QCD analysis of all available single inclusive unidentified charged hadron data sets in electron-positron ($e^+ e^-$) annihilations. 
The zero-mass variable flavor number scheme (ZM-VFNS) has been applied to account the heavy flavor contributions. The obtained optimum set of {\it residual} charged hadron FFs is accompanied by the well-known Hessian technique to assess the uncertainties in the extraction of these new sets of FFs. It is shown that the {\it residual} contributions of charged hadron FFs have very important impact on the inclusive charged hadron FFs and substantially on the quality and the reliability of the QCD fit. Furthermore, this study shows that the {\it residual} contributions become also sizable for the case of heavy quark FFs as well as for the $c$- and $b$-tagged cross sections.

\end{abstract}

\maketitle
\tableofcontents{}

%
\section{Introduction}\label{sec:introduction}
%

Quantum chromodynamics (QCD) is known as a fundamental theory of the strong interaction, and hence, it has been a topic of active research in the last decades~\cite{Gao:2017yyd,Ball:2017nwa}. Using the asymptotic freedom in QCD as well as in the perturbative theory, the high energy scattering processes can be analyzed. 
The factorization theorem separates the perturbative calculation part of the partonic cross section from the non-perturbative parts of both parton distribution functions (PDFs) and fragmentation functions (FFs). In hadronization processes, when specific hadrons are identified in the final state, FFs can explain how color-carrying quarks and gluons turn into the color-neutral hadrons. The common method to determine the FFs is to use the experimental data sets which are sensitive to the FFs. The well-known FFs are non-perturbative quantities in factorization theorem, and hence,  they need to be extracted from a global QCD analysis~\cite{Bertone:2018ecm,Nocera:2017gbk,dEnterria:2013sgr,deFlorian:2007ekg,Bourhis:2000gs,Kniehl:2000fe,Kretzer:2000yf,Aidala:2010bn,deFlorian:2017lwf,deFlorian:2014xna,Bertone:2017tyb,Sato:2016wqj,Hirai:2016loo,Anderle:2017cgl,Soleymaninia:2017xhc,Zarei:2015jvh,Boroun:2016zql,Helenius:2018uul}.

In the last decade, collinear or integrated FFs have been determined from neutral and charged hadrons in different high energy processes such as single inclusive electron-positron annihilation (SIA), semi-inclusive deep inelastic lepton-nucleon scattering (SIDIS) and single-inclusive hadron production in proton-proton collisions ~\cite{Bertone:2018ecm,Bertone:2017tyb,Soleymaninia:2018uiv,Soleymaninia:2017xhc,deFlorian:2014xna,deFlorian:2017lwf,Hirai:2016loo,Sato:2016wqj,Anderle:2017cgl}. Collinear or integrated FFs are denoted by $D^h_i(z)$ and describe the 
fragmentation of an unpolarized partons $f_i$ into unpolarized hadron $h$, where the fraction $z$ of the parton momentum is carried by the hadron. 
Since the FFs for different hadrons are universal quantities, they are calculated from different high energy processes at various center-of-mass energies.
The recent hadron production data sets  incorporate: The SIA measured by {\tt BELLE}~\cite{Leitgab:2013qh,Seidl:2015lla} and {\tt BaBar}~\cite{Lees:2013rqd}, semi-inclusive deep-inelastic scattering (SIDIS) measured by {\tt HERMES}~\cite{Airapetian:2012ki} and {\tt COMPASS}~\cite{Adolph:2016bga,Adolph:2016bwc} and, the (anti)proton collisions measured by {\tt CMS}~\cite{Chatrchyan:2011av,CMS:2012aa} and {\tt ALICE}~\cite{Abelev:2013ala} at the LHC, {\tt STAR}~\cite{Adamczyk:2013yvv} and {\tt PHENIX}~\cite{Adare:2007dg} at RHIC and {\tt CDF}~\cite{Abe:1988yu,Aaltonen:2009ne} at the Tevatron. These data sets cover a wide range of $\mu^2$ and $z$, and hence, they are sensitive to different parton species. Among them, the measurements of the $p_T$ charged-hadron spectra in proton-proton collisions are sensitive to the gluon FF, and therefore, provide the most stringent constraint on the gluon density~\cite{Bertone:2018ecm}.

Some of recent analyses determined different identified light charged hadrons, i.e. $\pi^\pm, K^\pm$ and $p/\bar{p}$~\cite{Bertone:2017tyb,deFlorian:2014xna,deFlorian:2017lwf} and unidentified light charge hadrons $h^\pm$~\cite{Bertone:2018ecm,Soleymaninia:2018uiv}. Recently, QCD analyses of heavier hadrons such as $D^*$ also have been done up to next-to-next-to leading order (NNLO)~\cite{Soleymaninia:2017xhc,Anderle:2017cgl}. Since the calculations for the hadronization processes in SIDIS and $pp$ collisions at NNLO are not yet accessible, these NNLO analyses are not global and only the SIA experimental data can be used in such analyses.
In recent years, there have been many studies to determine the unpolarized FFs for light or heavy hadrons. We refer the readers to the  Refs.~\cite{Bertone:2018ecm,Hirai:2016loo,deFlorian:2007aj,deFlorian:2007ekg,Soleymaninia:2013cxa} for more details. 
Although SIA experimental data provide the cleanest access to the FFs, and in comparison to SIDIS and $pp$ collision, the FFs are the only non-perturbative objects in the SIA cross sections, SIDIS data provides studying the flavor structure of FFs separately and $pp$ collision data is indispensable for constraining the gluon FFs. Recently, new analyses have been done to determine the unidentified light charged hadron FFs, and these FFs apply to the measurements of the charged-particle spectra in proton-ion and ion-ion collisions by RHIC~\cite{Adams:2005dq} and LHC~\cite{Abreu:2007kv}.  

In this paper, we determine the FFs of {\it residual} light charged hadrons at NLO and NNLO accuracies using the $e^+e^-$ annihilation experimental observables.  The unidentified light charged hadrons are considered as sum of the identified light charged hadrons such as pion, kaon, (anti)proton and the {\it residual} heavier charged hadrons. Hence, the {\it residual} light charged hadrons refers to the FFs for the fraction of hadrons that are not attributable to charged pions, charged kaons, protons or antiprotons. Although the contribution of the {\it residual} charged hadrons is small, but it is non-negligible.  Most recently, the NNPDF Collaboration extracted the FFs for unidentified charged hadrons and they have used the hadron production in proton-(anti)proton collisions data  as well as electron-positron annihilation data in their analysis entitled {\tt NNFF1.1h}~\cite{Bertone:2018ecm}. In addition, the {\tt NNFF1.0} have recently presented the FFs for charged pion, charged kaon and (anti)proton from an analysis of SIA hadron production data~\cite{Bertone:2017tyb}. Considering the pion, kaon and (anti)proton as the most important contributions in unidentified charged hadron production, we apply the FFs of {\tt NNFF1.0} analysis in order to determine the FFs of {\it residual} charged hadrons. Moreover, since the {\tt NNFF1.0} present the FFs of $\pi^\pm, K^\pm$ and $p(\bar{p})$ up to NNLO accuracy, it enable us to calculate the {\it residual} hadrons FFs up to NNLO approximation. We will show that the consideration of this small and important contributions of {\it residual} charged hadrons improves the agreements between theoretical predictions and the experimental data sets of unidentified charged hadron productions in SIA. Furthermore, this study shows that the {\it residual} contributions become also sizable for the case of heavy quark FFs as well as for the $c$- and $b$-tagged cross sections. 

The paper is organized as follows. In Sec.~\ref{sec:QCD analysis}, we discuss the perturbative QCD analysis of single-inclusive hadron production in electron-positron annihilation up to NNLO accuracy. Our methodology for the input parametrization at initial scale for the {\it residual} charged hadrons is presented in Sec.~\ref{sec:parametrization}. In Sec.~\ref{sec:data-selection}, we present all the experimental data sets analyzed in this study as well as the $\chi^2$ values calculated from our analyses for every data set. The minimization strategy to determine the FFs at initial scale and the Hessian uncertainty approach to calculate the errors of FFs are presented in Sec.~\ref{sec:errorcalculation}.
In Sec.~\ref{sec:Results}, we discuss the behavior of our FFs and compare them to other available FF sets in the literature.
We also present a detailed comparison of our theoretical predictions with the experimental data in this section. Finally, our summary and conclusion are given in Sec.~\ref{sec:conclusion}.

%
\section{ QCD analysis framework up to NNLO accuracy }\label{sec:QCD analysis}
%

In this section, we discuss in details the QCD analysis framework of FFs which is a well established perturbative QCD (pQCD) framework for analyzing the single-inclusive hadron production processes in $e^- e^+$ annihilation.
The cross section observables are defined based on the structure functions $F_{T, L, A}(z, \mu ^2)$ for the single inclusive $e^+e^-$ annihilation process of $e^+e^- \rightarrow \gamma/Z \rightarrow {\it h} + X $ at a given center-of-mass energy $\sqrt{s}$.
The general form for unpolarized inclusive single-particle production is  given by

\begin{eqnarray}\label{SIA}
\frac{1} {\sigma _0}
\frac{d^2 \sigma ^h}
{dz d\cos{\theta}}  & = &
\frac{3} {8}(1 + \cos ^2{\theta}) \
F_T^h(z, \mu ^2) + \frac{3}{4}
\sin ^2{\theta} \ F_L^h(z, \mu ^2)  \nonumber  \\
&+& \frac{3}{4} \cos {\theta} \
F^h_A(z, \mu ^2),
\end{eqnarray}

where $z=2E_h/{\sqrt{s}}$ is the scaled energy of the hadron $h$, and $\theta$ is the hadron angle relative to the electron beam. In above equation, the $F_T$ and $F_L$ are the transverse and longitudinal time-like structure functions, respectively.
The normalization factors $\sigma _0$ is equal to $\sigma _0=4\pi \alpha ^2 N_c/3s$.
Also the asymmetric structure function $F_A$ will be omitted by integration of Eq.~\eqref{SIA} over $\theta$, and hence, the total cross section can be written as

\begin{eqnarray}\label{totalcross}
&&\frac{1}{\sigma _{tot}}
\frac{d\sigma ^h}{dz} =
F_T^h(z, \mu ^2) + 
F_L^h(z, \mu ^2)   \nonumber  \\
&=& \sum _i \sum _a \int ^{1}_{z}
\frac{dx}{x}C_{i,a}(x, \alpha_s(\mu), 
\frac {s}{\mu ^2}) D^h_i(z/x, \mu^2)
+ {\cal O}(\frac{1} {\sqrt{s}}),  \nonumber \\
\end{eqnarray}

with $i=u, \bar{u}, d, \bar{d}, ... ,g$ and  $a=T$ and  $L$.
The differential cross section has been normalized to the total cross section
for $e^+e^-$ annihilation into hadrons $(\sigma_{tot})$
which reads $\sigma_{tot}=\sum _q e^2_q\sigma _{0}(1+\alpha_s(\mu ^2)/\pi$).

The function $D^h_i(z, \mu^2)$ is the fragmentation densities in which describe the probability that the parton $i$ fragments to a hadron $h$.
In above equation, $C_i$ are the process dependent coefficient functions which are given by

\begin{eqnarray}\label{wilson}
C_{a, i}(x, \alpha_s) = 
(1 - \delta_{aL}) 
\delta_{iq} +  
a_sc^{(1)}_{a, i}(x) + 
a_s^2c^{(2)}_{a, i}(x)
+ h.c. .
\end{eqnarray}

The coefficient functions are known up to NNLO approximation that have been reported in Refs.~\cite{Rijken:1996npa,Blumlein:2006rr,Mitov:2006wy}. According to Eq.~\eqref{wilson}, the coefficient functions for $F_L$ are vanished at leading order and the $C_L$ leading contribution is of order $\alpha_s$. The NNLO QCD corrections to the $F_L$ coefficient functions, which are ${\cal O}(a^3_s)$, are not known in the literature. Since the perturbative corrections to the coefficient functions of the longitudinal cross section are only known up to ${\cal O} (\alpha_s^2)$, one cannot analyses the longitudinal structure function $F_L$ data at NNLO accuracy. In addition, we use the publicly available {\tt APFEL} code~\cite{Bertone:2013vaa} to perform our analysis and the NNLO QCD corrections to the $F_L$ are not included in this code. Since NNLO QCD corrections to the corresponding coefficient functions, which are $O(\alpha^3_s)$, are not known in the literature nor in the {\tt APFEL} code, we do not include the longitudinal experimental data sets in our analysis.

Perturbative QCD corrections lead to logarithmic scaling violations via the DGLAP evolution equations~\cite{Gribov:1972ri,Lipatov:1974qm,Altarelli:1977zs,Dokshitzer:1977sg} which evaluate the FFs with the energy scale $Q^2$ as

\begin{eqnarray}\label{non-singlet DGLAP}
\frac{\partial}
{\partial \ln \mu^2} D_i(z, \mu^2) =
\sum_{j}\int ^1_z\frac{dx} {x} P_{ji}(x, \alpha_s(\mu^2)) 
D_{j}(\frac{z} {x}, \mu^2) \,
\end{eqnarray}

where $P_{ji}(x,\alpha_s(\mu^2))$ are  purturbative splitting functions

\begin{eqnarray}\label{split}
P_{j,i}(x, \alpha_{s}) = 
\frac{\alpha_s}{2\pi}
P_{ji}^{(0)}(x) +
(\frac{\alpha_{s}}{2 \pi})^2 P_{ji}^{(1)}(x)
+ (\frac{\alpha_{s}}{2\pi})^3 P_{ji}^{(2)} (x)
+  h.c. .
\end{eqnarray}

Commonly, the DGLAP equation is decomposed into a $2 \times2$ flavour-singlet sector comprising the sum
of all quark and antiquark fragmentation functions and gluon as well as the non-singlet equations for quark-antiquark and flavour differences. The evolution of the FFs $D_i(z, \mu^2)$ in Eq.~\eqref{non-singlet DGLAP} as well as the numerical computation of the cross section in Eq.~\eqref{totalcross} are performed using the publicly available {\tt APFEL} package \cite{Bertone:2015cwa,Bertone:2013vaa} at NLO and NNLO accuracy in pQCD. This package has been used in our pioneering works in Refs.~\cite{Soleymaninia:2018uiv,Soleymaninia:2017xhc} as well as many other analyses in literature such as {\tt NNFF}~\cite{Bertone:2018ecm,Nocera:2017gbk,Bertone:2017tyb}.

%
\section{ Phenomenological parametrization up to NNLO } \label{sec:parametrization}
%

In this section, we will describe all techniques including the phenomenological parametrization as well as the assumptions we use for the global analysis of {\it residual } charged hadrons FFs. The unidentified charged hadrons are sum of the identified light charged hadrons that are produced in the fragmentation of a parton. The light charged hadrons include pion ($\pi ^\pm$), kaon ($K^\pm$), proton ($p/\bar{p}$) and {\it residual} light hadrons. Then, the unidentified charged hadron cross sections can be written as a sum of the individual cross sections of $\pi ^\pm$, $K^\pm$, $p/\bar{p}$ and {\it residual} hadrons.  Following that, the unidentified charged hadron FFs are sum of the FFs of $\pi ^\pm$, $K^\pm$, $p/\bar{p}$ and {\it residual} light hadrons,

\begin{eqnarray}\label{residual_1}
D_i^{h^\pm}(z, \mu^2) =  D_i^{\pi^\pm} (z, \mu^2) +  D_i^{K^\pm} (z, \mu^2)  +  D_i^{p/\bar{p}} (z, \mu^2)  + D_i^{{\it res}^\pm} (z, \mu^2).
\end{eqnarray}

Consequently, in order to calculate the FFs of {\it residual} hadrons we use the following relation,

\begin{eqnarray}\label{residual}
D_i^{{\it res}^\pm}(z, \mu^2) = D_i^{h^\pm}(z, \mu^2)-\sum_l D_i^{l} (z, \mu^2), ~~~ l= \pi^\pm, K^\pm, p/\bar{p}.
\end{eqnarray}

Our main aim in this analysis is the determination of $D^{{res}^\pm}(z, \mu^2)$ by including SIA experimental data of the unidentified light charged hadrons and also using the FFs of charged pions, charged kaons, and (anti)protons from the recent {\tt NNFF1.0} sets~\cite{Bertone:2017tyb}. The FFs of {\tt NNFF1.0} have been determined from an analysis of single inclusive hadron production data in electron-positron annihilation at leading order (LO), NLO and NNLO accuracy.

In the following, we introduce the methodology and the assumptions of our analysis to determine the {\it residual} charged hadrons FFs. In comparison to the other light hadrons, pion productions are much more copiously and after pions the production of kaons and protons are more than the others. Then we expect that the $D^{h^\pm}$ is strongly dominated by these three light hadrons and then the contribution of {\it residual} light hadrons in Eq.~\eqref{residual} seems to be small but rather important. 
Hence, we choose the most simple functional form for all the parton FFs as follows,

\begin{eqnarray}\label{parametrization}
D^{{\it res}^\pm}_i(z, \mu_0^2) = \frac{ {N} _i
z^{\alpha_i}(1 - z)^{\beta_i}} {B[2 + \alpha_i,
\beta_i + 1]}, ~~~ i = u + \bar{u}, d + \bar{d}, s + \bar{s}, c + \bar{c}, b + \bar{b}, g.
\end{eqnarray}

The $N_i$ in above equation represents the normalizations of FFs and along with the free parameters $\{\xi_i = \alpha_i, \, \beta_i \}$, they need to be determined from QCD fit to the data. The variation of the {\it residual} light hadrons distributions at small and large values of momentum fraction $z$ will be controlled by the ${\alpha_i}$ and ${\beta_i}$, respectively.

The extraction of charged hadrons FFs in a global QCD analysis of a large body of data at NLO as well as NNLO accuracy requires an extensive number of
time-consuming computations of the FFs evolution and the corresponding observables in each step of the usual $\chi^2$ minimization procedure.
The large number of parameters specifying the functional form of the charged hadrons FFs in the QCD fit
and the need
for a proper assessment of their uncertainties, add to this. Hence, we prefer to choose a simple standard form for our {\it residual } charged hadrons FFs as presented in Eq.~\eqref{parametrization}.
In addition, the available SIA data are not accurate enough to determine all the shape parameters with sufficient accuracy, and hence, it encourages us to assume a very simple form for the {\it residual } charged hadrons FFs.
	
It should be noted that in our analysis, the initial scale for the above parametrization form is $\mu _0=5$~GeV for all partons. Since we use very recent {\tt NNFF1.0} sets for $\pi ^\pm, K^\pm$ and $p/\bar{p}$, we choose the {\tt NNFF1.0} initial scale in our analysis. Also the value of charm and bottom masses in our analysis are same as the {\tt NNFF1.0}, and hence, we fixed them to $m_c=1.51$ and $m_b=4.92$.  We should emphasize here that we use the fragmentation functions of pions, kaons, and protons/antiprotons from {\tt NNFF1.0} set~\cite{Bertone:2017tyb} at the input parametrization scale $\mu_0$ = 5 GeV, and then we evolved these FFs with our {\it residual} charged hadrons FFs using the {\tt APFEL} kernel~\cite{Bertone:2013vaa}.

Let us now discuss our final definitions of the {\it residual } charged hadrons FFs considered in this analysis. As a first assumption, the $SU(3)$ flavor symmetry is considered for the  light quarks $(u,d,s)$ since the data are not sensitive to the kind of light quarks, such that
\begin{eqnarray}\label{symmetruy}
D^{{\it res}^\pm}_{u+\bar{u}} = D^{{\it res}^\pm}_{d+\bar{d}}= D^{{res}^\pm}_{s+\bar{s}} \,.
\end{eqnarray}
We should mentioned that the charge conjugation symmetry is another assumption that we considered in our analysis, i.e., $D^{{res}^+}_{q} =D^{{res}^-}_{\bar{q}}$.  As we previously discussed, based on the SIA tagged data sets we included, they are only sensitive to the flavor combinations of $u + \bar{u} + d + \bar{d} + s + \bar{s}$, $c + \bar{c}$ and $b + \bar{b}$. Then we can choose separate parametrization form for the heavy quark FFs. 
 Hence, in total, we have 12 free parameters in our parametrization for the {\it residual } charged hadrons FFs.  We should highlight here that, during the fit procedure and constraining the fit parameters, we found  that the data used in this analysis can not really put enough constrain for all the parameters and then some of the parameters should be fixed in the best values of the first minimization. Hence, we fix two of parameters, namely $\alpha_{u + \bar{u}}^{\rm NLO} = 154.59$, $\alpha^{\rm NLO}_g = 27.36$, $\alpha_{u + \bar{u}}^{\rm NNLO} = 153.47$ and $\alpha^{\rm NNLO}_g = 24.08$ at their best fit values at NLO and NNLO accuracy (see Tables.~\ref{tab:NLO} and \ref{tab:NNLO} for more details.). Finally, these assumptions lead to $10$ free parameters which should be extracted from the QCD fit to data to determine the FFs uncertainties.

In the forthcoming sections, we discuss the data sets included in this analysis and then we present the $\chi^2$ function and the various methods
for the analysis of {\it residual } charged hadrons FFs uncertainties. Most of the discussions presented here will follow the pioneering work in Ref.~\cite{Soleymaninia:2018uiv}.

%
\section{ Description of experimental observables } \label{sec:data-selection}
%

In this section, we will review the experimental data sets used in this analysis to determine the {\it residual}  charged hadrons FFs.
As we mentioned earlier, we restrict this analysis to SIA and consider all available tagged and flavor-untagged resentments performed by different experiments, including {\tt ALEPH}, {\tt  OPAL} and {\tt  DELPHI} experiments at CERN, {\tt TASSO} experiment at DESY, and {\tt TPC} and {\tt SLD} experiments at SLAC. 
The analyzed SIA data sets are summarized in Tables.~\ref{tab:datasets-TASSO}, \ref{tab:datasets-TPC}, \ref{tab:datasets-ALEPH}, \ref{tab:datasets-DELPHI}, \ref{tab:datasets-OPAL} and \ref{tab:datasets-SLD}. For each data sets we specify the name of the experiment, the corresponding reference, the observable, the center-of-mass energy $\sqrt{s}$ and the number of analyzed data points for each experiment. These tables also include the $\chi^2$ values for both NLO and NNLO analyses. As we mentioned, SIA data sets are fundamental quantities providing information about quark fragmentation and are also sensitive to the flavor of $q + \bar q$ fragmentation functions. 

In order to avoid the resummation effects at small and large $z$ regions, we exclude the data in theses regions. According to the reasonable result in our analysis, we choose the value $z_{min}=0.02$ for data sets at $\mu=M_Z$ and $z_{min}=0.075$ for $\mu <M_Z$. The kinematic cut for large $z$ is taken to be $z_{max}=0.9$ for all data sets in our fits. These selections on SIA data sets are same as recent analyses by {\tt NNFF} collaboration~\cite{Bertone:2018ecm,Nocera:2017gbk}. Considering the kinematic cuts, we include the total 474 data points at both NLO and NNLO QCD fits. 
It should be mentioned here that since we include pion, kaon, and proton FFs from {NNFF1.0} analysis, their uncertainties should be considered in the theoretical calculations of the unidentified charged hadron cross sections.
For the uncertainty from {\tt NNFF1.0}, we follow the analysis by {\tt DSS07} in Ref.~\cite{deFlorian:2007ekg} and estimate an average uncertainty of 5\% in all theoretical calculations of the inclusive charged hadron cross sections stemming from the uncertainties of pion, kaon, and proton FFs from {\tt NNFF1.0} set. This additional uncertainty is included in the $\chi^2$ minimization procedure for determining the {\it residual} charged hadrons FFs. We apply the simplest way to include a ``theory'' error which is to add it in quadrature to the statistical and systematic experimental error in the $\chi^2$ expression.  We should mentioned here that the uncertainties from {\tt NNFF1.0} parameterizations are not flat over $z$ and also depend on this variable, hence one need to properly propagate these uncertainties through the QCD analysis. However, like for the case of {\tt DSS07} analysis, we believe that a 5\% of the cross section value seems to be reasonable. 

In the following, we begin with discussing the measurements of single-inclusive charged hadron production in electron-positron annihilation, collected by different experiments. 
The first source of information on the unidentified charged hadrons is provided by {\tt TASSO} experiment at DESY for the total inclusive cross section measurements for annihilation into hadron according to the reaction $e^+ e^- \to {\rm hadrons}$~\cite{Braunschweig:1990yd}. As indicated in Table.~\ref{tab:datasets-TASSO}, these data sets correspond to the four different center-of-mass energies of $\sqrt{s}$ = 14, 22, 35 and 44 GeV. This measurement covers the range of $14 \leq Q \leq 44$ GeV and $0.025 \leq z \leq 0.9$.
After applying kinematical cuts on the analyzed data sets, we use 60 data points from {\tt TASSO} experiment.

%
\begin{table}[htb]
\renewcommand{\arraystretch}{2}
\centering 	\scriptsize
\begin{tabular}{lc||l|cc||cc|cc}
\toprule \hline \hline
{\tt Experiment} ~&~ {\tt Reference} ~&~ {\tt Observable} ~&~  $\sqrt{s}$~[{\tt  GeV}] ~&~  {\tt Number of data points} ~&~  ${\cal N}^{{\tt NLO}}$ ~&~ $\chi_{n}^{2,{\tt NLO}}$ ~&~ ${\cal N}^{{\tt NNLO}}$ ~&~ $\chi_{n}^{2,{\tt NNLO}}$ \\ \hline \hline
\midrule 
{\tt TASSO-14} & \cite{Braunschweig:1990yd} & $\frac{1} {\sigma_{\mathrm{total}}}~\frac{d\sigma^{{\mathrm h}^\pm}} {dz}$ 		& 14.00  		& ~~~~~15  		& 1.0108 & 8.83 & 1.0082 & 7.75 \\
{\tt TASSO-22} & \cite{Braunschweig:1990yd}
& $\frac{1} {\sigma_{\mathrm{total}}}~\frac{d \sigma^{{\mathrm h}^\pm}} {dz}$ 		& 22.00  		& ~~~~~15   	& 1.0127 & 15.13   & 1.0111 & 14.71  \\
{\tt TASSO-35} & \cite{Braunschweig:1990yd}
& $\frac{1} {\sigma_{\mathrm{total}}}~\frac{d\sigma^{{\mathrm h}^\pm}} {dz}$ 		& 35.00  		& ~~~~~15     & 1.0135	& 29.92  & 1.0102 & 31.48   \\ 
{\tt TASSO-44} & \cite{Braunschweig:1990yd}
& $\frac{1} {\sigma_{\mathrm{total}}}~\frac{d\sigma^{{\mathrm h}^\pm}} {dz}$ 		& 44.00  		& ~~~~~15  		& 1.0146 & 16.86  & 1.0125 & 16.52  \\ \hline \hline \midrule
\bottomrule
\end{tabular}
\caption{ \small The data sets by {\tt TASSO} experiment at DESY used in the present analysis of FFs for {\it residual} charged hadrons.
For each experiment, we present the observables and corresponding reference, the center-of-mass energy $\sqrt{s}$, the number of analyzed data points 
after kinematical cuts, and the $\chi^2$ values for each data set. The details of corrections and the kinematical cuts applied are contained in the text. }
\label{tab:datasets-TASSO}
\end{table}

The total inclusive cross section measurements by {\tt TPC} experiment at SLAC~\cite{Aihara:1988su} for unidentified charged hadrons is presented in Table.~\ref{tab:datasets-TPC}. This data correspond to the center-of-mass energy of $\sqrt{s}$ = 29 GeV for the momentum interval $0.01 \leq z \leq 0.9$. 

%
\begin{table}[htb]
\renewcommand{\arraystretch}{2}
\centering 	\scriptsize
\begin{tabular}{lc||l|cc||cc|cc}
\toprule \hline \hline
{\tt Experiment} ~&~ {\tt Reference} ~&~ {\tt Observable} ~&~  $\sqrt{s}$~[{\tt  GeV}] ~&~  {\tt Number of data points} ~&~  ${\cal N}^{{\tt NLO}}$ ~&~ $\chi_{n}^{2,{\tt NLO}}$ ~&~ ${\cal N}^{{\tt NNLO}}$ ~&~ $\chi_{n}^{2,{\tt NNLO}}$ \\ \hline \hline
\midrule 
{\tt TPC} & \cite{Aihara:1988su}
& $\frac{1} {\sigma_{\rm{total}}}~\frac{d\sigma^{{\it h}^\pm}} {dz}$ 		& 29.00  		& ~~~~~21 		& 1.0331 & 32.86 & 1.0304 &  30.03   \\ \hline \hline		\midrule
\bottomrule
\end{tabular}
\caption{ \small The data sets by {\tt TPC} experiment at SLAC used in the present analysis of FFs for {\it residual} charged hadrons.  See the caption of Table.~\ref{tab:datasets-TASSO} for further details. }
\label{tab:datasets-TPC}
\end{table}

Another source of information for unidentified charged hadrons comes from the data collected by {\tt ALEPH} experiment at CERN~\cite{Buskulic:1995aw}.
As one can see from Table.~\ref{tab:datasets-ALEPH}, these data sets correspond to the totalinclusive cross section measurements of charged particles for the center-of-mass energy of $\sqrt{s} = M_Z$.
%
\begin{table}[htb]
\renewcommand{\arraystretch}{2}
\centering 	\scriptsize
\begin{tabular}{lc||l|cc||cc|cc}
\toprule \hline \hline
{\tt Experiment} ~&~ {\tt Reference} ~&~ {\tt Observable} ~&~  $\sqrt{s}$~[{\tt  GeV}] ~&~  {\tt Number of data points} ~&~  ${\cal N}^{{\tt NLO}}$ ~&~ $\chi_{n}^{2,{\tt NLO}}$ ~&~ ${\cal N}^{{\tt NNLO}}$ ~&~ $\chi_{n}^{2,{\tt NNLO}}$ \\ \hline \hline
\midrule 
{\tt ALEPH} & \cite{Buskulic:1995aw}
& $\frac{1} {\sigma_{\rm{total}}}~\frac{d\sigma^{{\it h}^\pm}} {dz}$ 		& 91.20  		& ~~~~~32  		& 0.9997 & 1.89 & 0.9999 & 2.01  
\\   \hline \hline		\midrule
\bottomrule
\end{tabular}
\caption{ \small The data sets by {\tt ALEPH} experiment at CERN used in the present analysis of FFs for {\it residual} charged hadrons. See the caption of Table.~\ref{tab:datasets-TASSO} for further details. }
\label{tab:datasets-ALEPH}
\end{table}
%

In Tables.~\ref{tab:datasets-DELPHI} and \ref{tab:datasets-OPAL}, we indicate another key ingredient in our {\it residual} charged hadrons FFs analysis which are the single inclusive hadron production data sets from electron-positron collisions at {\tt  DELPHI} and {\tt OPAL} experiments at CERN~\cite{Abreu:1998vq,Abreu:1997ir,Ackerstaff:1998hz}. 

%
\begin{table}[htb]
\renewcommand{\arraystretch}{2}
\centering 	\scriptsize
\begin{tabular}{lc||l|cc||cc|cc}
\toprule \hline \hline
{\tt Experiment} ~&~ {\tt Reference} ~&~ {\tt Observable} ~&~  $\sqrt{s}$~[{\tt  GeV}] ~&~  {\tt Number of data points} ~&~  ${\cal N}^{{\tt NLO}}$ ~&~ $\chi_{n}^{2,{\tt NLO}}$ ~&~ ${\cal N}^{{\tt NNLO}}$ ~&~ $\chi_{n}^{2,{\tt NNLO}}$ \\ \hline \hline
\midrule 
{\tt DELPHI} & \cite{Abreu:1998vq}   
& $\frac{1} {\sigma_{\rm{total}}}~\frac{d\sigma^{{\it h}^\pm}}{dp_{\it h}}$ 		& 91.20  		& ~~~~~22  		& 0.9965 & 10.94   & 0.9972 & 8.81  \\
& \cite{Abreu:1998vq}
& $\left.\frac{1} {\sigma_{\rm{total}}}~\frac{d\sigma^{{\it h}^\pm}}{dp_{\it h}}\right |_{uds}$ 		& 91.20  		& ~~~~~22  		& 1.0046 & 10.03  & 1.0051 & 8.04 \\
& \cite{Abreu:1998vq}
& $\left.\frac{1} {\sigma_{\rm{total}}}~\frac{d\sigma^{{\it h}^\pm}}{dp_{\it h}}\right |_{b}$ 		& 91.20  		& ~~~~~22  		& 0.9902 &  11.48   & 0.9932 & 10.34    \\  \hline \hline	\midrule
\bottomrule
\end{tabular}
\caption{ \small The data sets by {\tt DELPHI} experiment at CERN used in the present analysis of FFs for {\it residual} charged hadrons. See the caption of Table.~\ref{tab:datasets-TASSO} for further details. }
\label{tab:datasets-DELPHI}
\end{table}
%

%
%
\begin{table}[htb]
\renewcommand{\arraystretch}{2}
\centering 	\scriptsize
\begin{tabular}{lc||l|cc||cc|cc}
\toprule \hline \hline
{\tt Experiment} ~&~ {\tt Reference} ~&~ {\tt Observable} ~&~  $\sqrt{s}$~[{\tt  GeV}] ~&~  {\tt Number of data points} ~&~  ${\cal N}^{{\tt NLO}}$ ~&~ $\chi_{n}^{2,{\tt NLO}}$ ~&~ ${\cal N}^{{\tt NNLO}}$ ~&~ $\chi_{n}^{2,{\tt NNLO}}$ \\ \hline \hline
\midrule 
{\tt OPAL} & \cite{Ackerstaff:1998hz}
& $\frac{1} {\sigma_{\rm{total}}}~\frac{d\sigma^{{\it h}^\pm}} {dz}$ 		& 91.20  		& ~~~~~20   		& 0.9915 & 6.30 & 0.9934 &  5.40  \\ 
& \cite{Ackerstaff:1998hz}
& $\left.\frac{1} {\sigma_{\rm{total}}}~\frac{d\sigma^{{\it h}^\pm}} {dz}\right |_{uds}$ 		& 91.20  		& ~~~~~20   		& 0.9991 & 10.04  & 0.9998 & 8.92  \\ 
& \cite{Ackerstaff:1998hz}
& $\left.\frac{1} {\sigma_{\rm{total}}}~\frac{d\sigma^{{\it h}^\pm}} {dz}\right |_{c}$ 		& 91.20  		& ~~~~~20  		& 0.9936 & 16.31  & 0.9939 &  16.34  \\ 
& \cite{Ackerstaff:1998hz}
& $\left.\frac{1} {\sigma_{\rm{total}}}~\frac{d\sigma^{{\it h}^\pm}} {dz}\right |_{b}$ 		& 91.20  		& ~~~~~20  		& 0.9951 & 4.86  & 0.9982 &  4.13  \\   \hline \hline \midrule
\bottomrule
\end{tabular}
\caption{ \small The data sets by {\tt OPAL} experiment at CERN used in the present analysis of FFs for {\it residual} charged hadrons. See the caption of Table.~\ref{tab:datasets-TASSO} for further details. }
\label{tab:datasets-OPAL}
\end{table}

Finally, the last source of information on the unidentified charged hadrons is provided by the {\tt SLD} experiments at SLAC, (see Table.~\ref{tab:datasets-SLD}). {\tt SLD} data sets correspond to the center-of-mass energy of $\sqrt{s} = 91.28$ GeV~\cite{Abe:2003iy}. In total, after kinematic cuts, we use 136 data points provided by this experiment.

%
\begin{table}[htb]
\renewcommand{\arraystretch}{2}
\centering 	\scriptsize
\begin{tabular}{lc||l|cc||cc|cc}
\toprule \hline \hline
{\tt Experiment} ~&~ {\tt Reference} ~&~ {\tt Observable} ~&~  $\sqrt{s}$~[{\tt  GeV}] ~&~  {\tt Number of data points} ~&~  ${\cal N}^{{\tt NLO}}$ ~&~ $\chi_{n}^{2,{\tt NLO}}$ ~&~ ${\cal N}^{{\tt NNLO}}$ ~&~ $\chi_{n}^{2,{\tt NNLO}}$ \\ \hline \hline
\midrule 
{\tt SLD} & \cite{Abe:2003iy}
& $\frac{1} {\sigma_{\rm{total}}}~\frac{d\sigma^{{\it h}^\pm}}{dp_h}$ 	                     	& 91.28  		& ~~~~~34 		& 1.0006 & 28.70 & 1.0007 &  21.83  \\ 
& \cite{Abe:2003iy}
& $\left. \frac{1} {\sigma_{\rm{total}}}~\frac{d\sigma^{{\it h}^\pm}}{dz} \right |_{uds}$ 		& 91.28   		& ~~~~~34 		& 1.0006 & 18.38 & 1.0008 & 17.57  \\ 
& \cite{Abe:2003iy}
& $\left. \frac{1} {\sigma_{\rm{total}}}~\frac{d\sigma^{{\it h}^\pm}}{dz} \right |_{c}$ 		& 91.28   		& ~~~~~34  		& 0.9995 & 38.88 & 0.9994 & 36.03  \\ 
& \cite{Abe:2003iy}
& $\left. \frac{1} {\sigma_{\rm{total}}}~\frac{d\sigma^{{\it h}^\pm}}{dz} \right |_{b}$ 		& 91.28  		& ~~~~~34 		& 1.0002 & 7.70  & 1.0004 &  7.86  \\ \hline \hline \midrule
\bottomrule
\end{tabular}
\caption{ \small The data sets by {\tt SLD} experiment at CERN used in the present analysis of FFs for {\it residual} charged hadrons. See the caption of Table.~\ref{tab:datasets-TASSO} for further details. }
\label{tab:datasets-SLD}
\end{table}

As one can see from the experiments outlined in this section, variety of SIA data sets have been used in our analysis to extract the {\it residual} charged hadrons FFs. The flavor tagged cross sections, could help to distinguish between the sum of light $u$, $d$ and $s$-quarks, as well as the charm and bottom FFs. 

Stringent constraint for bottom FF comes mainly from the {\tt DELPHI}, {\tt OPAL} and {\tt SLD} flavor tagged data sets. For charm FF, {\tt OPAL} and {\tt SLD} data sets with slightly larger errors are available. As one can expect, the singlet combination $q + \bar{q}$ is constrained well enough with the electron-positron annihilation data, while for gluon FF these data sets could not provide enough information. Gluon FF is constrained in our fit by the scale dependence of the data. The measurements of the longitudinal inclusive cross sections can be used to extract the fragmentation function for the gluon. Due to the lack of precise data that cover a wide range of energies, this data could also helps to constrain the gluon FF. The motivation for using the  $c$-tagged and $b$-tagged data in our analysis comes mainly from the ability to separate the heavy flavor FFs for charm and bottom. We show that all the analyzed data sets are reasonably well described by our QCD fits.

In the next section, we present the calculation method of uncertainties for the resulting new set of {\it residual } charged hadrons FFs.

%
\subsection{The minimization of {\it residual } charged hadrons FFs } \label{sec:errorcalculation}
%

To determine the best values of the known parameters at NLO and NNLO accuracies, one need to minimize the $\chi^2$ with respect to four free input {\it residual } charged hadrons FFs parameters of Eqs.~(\ref{parametrization}). 
In a global QCD analyses of PDFs as well as FFs, the global goodness-of-fit procedure usually follows the usual method with $\chi^2 (\{\xi\})$ defined as

\begin{equation}\label{chi2}
\chi^2(\{\xi\}) = \sum_{i = 1}^{n^{\tt{data} }}
\frac{ (~ {\cal D}_i^{\tt{data} }
- {\cal T}_i^{\tt{theory} } (\{\xi\}) ~ )^2}
{ ( \sigma_i^{\tt{data} } )^2 } \,,
\end{equation}

where $\{\xi\}$ denotes the set of independent free parameters in the fit, and $n^{\tt data}$ is the number of data points included in this analysis, which is $n^{\tt data}$ = 474. 
In Eq.~\eqref{chi2}, the quantity ${\cal D}^{\tt{data}}$ is the measured value of a given observable and ${\cal T}^{\tt{theory}}$ is the corresponding theoretical estimate for a
given set of parameters $\{\xi\}$ at the same experimental $z$ and $Q^2$ points. The widely-used CERN program library {\tt MINUIT}~\cite{CERN-Minuit} is applied to obtain the best parametrization of the  {\it residual } charged hadrons FFs. The experimental errors are calculated from systematic and statistical errors added in quadrature, $(\sigma_i^{\tt{data}})^2 = (\sigma_i^{\tt{sys }})^2 + (\sigma_i^{\tt{stat}})^2$.
	
For all analyzed data sets, we obtained $\chi^2/{\tt dof} = 0.699$ for the NLO analysis and $\chi^2/{\tt dof} = 0.643$ for the NNLO one, which indicate that the inclusion of higher order corrections lead to  improvement in the $\chi^2/{\tt dof}$. 
The $\chi^2$ values corresponding to each individual data set for each of the NLO and NNLO fits are presented in Tables.~\ref{tab:datasets-TASSO}, \ref{tab:datasets-TPC},  \ref{tab:datasets-ALEPH},  \ref{tab:datasets-DELPHI}, \ref{tab:datasets-OPAL} and \ref{tab:datasets-SLD} for the {\tt TASSO}, {\tt TPC}, {\tt ALEPH}, {\tt DELPHI}, {\tt OPAL} and {\tt SLD}, respectively. As one can see, almost for all data set, the NNLO QCD correction lead to the reduction of individual $\chi^2$. 
 
As one can see, for some certain experiments such as {\tt TASSO-35} we obtained relatively large value of $\chi^2$ showing a lower agreement in comparison with the other datasets between our theory predictions and this particular set of data. For other experiment such as {\tt ALEPH}, the $\chi^2$ is slightly too small. These treatments may deserve some detailed discussions. By refereeing to the analysis by DSS07~\cite{deFlorian:2007ekg} in which residual unidentified light charged hadron is determined at NLO, one can see the same conclusion in their analysis. They obtained a relatively small value of $\chi^2$ for the {\tt TPC} and large value for the DELPHI. It should be emphasize now that the kinematical cuts for the $z$ in DSS07 analysis are different with the cuts  we applied in our analysis. They excluded the data sets in the $z \le 0.1$ region while we choose the value $z_{min} = 0.02$ for data sets at $Q=M_Z$ and $z_{min} = 0.075$ for $Q<M_Z$.

We should mention here that most single-inclusive charged hadron production data in electron-positron annihilation come with an additional information on the fully correlated normalization uncertainty. Since, the simple $\chi^2(\{\xi\})$ definition needs to be modified in order to account for such normalization uncertainties. Therefore, the modified $\chi^2$ function is given by:

\begin{eqnarray}\label{eq-chi2global}
\chi_{\tt global}^2 (\{\xi_{i}\}) & = &
\sum_{n = 1}^{n^{\tt exp}} \left( \frac{1 - {\cal N}_n }
 {\Delta{\cal N}_n}\right)^2  + \nonumber \\ 
&&\sum_{j = 1}^{{\tt N}_n^{\tt data}}
 \left(\frac{ ( {\cal N}_n
\, {\cal D}_{j}^{\tt data} - {\cal T}_{j}^{\tt theory}
 (\{\xi_i\})}
{{\cal N}_n \, \delta {\cal D}_{j}^{\tt data}}
 \right)^2 \,,   \nonumber \\ 
\end{eqnarray}

$\Delta{\cal N}_n$ in above equation are the experimental normalization uncertainties quoted by the experiments. The relative normalization factors ${\cal N}$ can be fitted along with the fitted parameters $\{\xi\}$ of  {\it residual } charged hadrons FFs and then kept fixed. The relative normalization factors for our NLO (${\cal N}^{{\tt NLO}}$)  and NNLO  (${\cal N}^{{\tt NNLO}}$)  analyses extracted from fit to the data, are presented in Tables.~\ref{tab:datasets-TASSO}, \ref{tab:datasets-TPC},  \ref{tab:datasets-ALEPH},  \ref{tab:datasets-DELPHI},  \ref{tab:datasets-OPAL} and \ref{tab:datasets-SLD}.

%
\subsection{ Uncertainties of {\it residual } charged hadrons FFs }
%

An important objective in a global QCD analysis of FFs is to estimate uncertainties of the charged hadrons FFs obtained from the $\chi^2$ optimization.
To this end, in the following section, we present our method for the calculation of the {\it residual } charged hadrons FFs uncertainties and error propagation from experimental data points. To obtain the uncertainties in any global FFs analyses, there are well-defined procedures for propagating experimental uncertainties on the fitted data points through to the FFs uncertainties. In this paper, ``Hessian method'' will mainly be the method of our choice for estimating uncertainties of the {\it residual } charged hadrons FFs.
Hence, in our analysis, we apply the ``Hessian method'' (or error matrix approach), which is based on linear error propagation and involves the production of eigenvector FFs sets suitable for convenient use by the end user.

Originally, the ``Hessian method'' was widely used in MRST~\cite{Martin:2003sk} and MSTW08~\cite{Martin:2009iq} global QCD analyses and we also applied this approach in our previous works~\cite{AtashbarTehrani:2012xh,Khanpour:2016pph,Shahri:2016uzl}. Therefore, in the present analysis,  we again follow this method and extract the uncertainties of {\it residual } charged hadrons FFs. Following that, an error analysis can be obtained by using the ``Hessian matrix'', which is determined by running the CERN program library MINUIT~\cite{CERN-Minuit}.

The most commonly applied Hessian approach, which is based on the covariance matrix diagonalization, provides us a simple and
efficient method for calculating the uncertainties of {\it residual } charged hadrons FFs.
The basic assumption of the Hessian approach is a quadratic expansion of the global goodness-of-fit quantity, $\chi^2_{\text{ global}}$, in the fit parameters $\xi_i$ near the global minimum,

\begin{equation} \label{chi2-2}
\Delta \chi^2_{\tt{ global}} \equiv  \chi^2_{\tt{global}}  - \chi^2_{\tt min} = \sum_{i, j=1}^n (\xi_i - \xi_i^0) \, H_{ij} \,   (\xi_j - \xi_j^0) \,,
\end{equation}

where $H_{\rm ij}$ are the elements of the Hessian matrix and $n$ stands for the  number of parameters in the global fit.

The uncertainty on a {\it residual } charged hadrons FFs $D^{\tt res}(z, \xi_i)$ is then given by

\begin{eqnarray}\label{deltak2}
\delta D^{\tt res}(z, \xi_i)& = &  \nonumber \\
\biggl[\Delta \chi^2 && \sum_{i, j}^n 
\left ( \frac{\partial D^{\tt res}(z, \xi)}
{\partial \xi_i}
\right) _{\xi = \hat{\xi}}  H_{ij}^{-1}
\left ( \frac{\partial D^{\tt res}(z, \xi)}
{\partial \xi_j} \right)_{\xi = \hat{\xi}} 
\biggr]^{1/2}\,, \nonumber \\
\end{eqnarray}

where $\xi_i$ stand for the fit parameters in the input  {\it residual } charged hadrons FFs, and $\hat {\xi}$ indicates the number of parameters which make an extreme value for the related derivative. Running the CERN program library {\tt MINUIT}, the Hessian or covariance matrix elements for free parameters in our NLO and NNLO {\it residual } charged hadrons FFs analyses can be obtained. The uncertainties of {\it residual} charged hadrons FFs as well as the related observable are estimated using the ``Hessian matrix'' explained above and their values at higher $\mu^2$ ($\mu^2  > \mu_0^2$) are calculated using the DGLAP evolution equations.

%
\section{ Discussion of QCD fit results and {\it residual} charged hadrons FFs } \label{sec:Results}
%

Now we turn to the numerical results for the {\it residual} charged hadrons FFs extracted from the following analyses at NLO and NNLO accuracy. In Tables.~\ref{tab:NLO} and \ref{tab:NNLO} we present the best fit parameters for the fragmentation of quarks and gluon into the $D^{{res}^\pm}$ at NLO and NNLO accuracy in pQCD. As we mentioned before, the starting scale is taken to be $Q_{0}=5$~GeV for all parton species. The values labeled by (*) have been fixed after the first minimization, since the analyzed SIA data dose not constrain all unknown fit parameters well enough.  Regarding the simple parameterization that we considered in this analysis, one can see from Tables.~\ref{tab:NLO} and \ref{tab:NNLO} that we fixed the parameter $\alpha$ for the $u+ \bar u$ and gluon FFs. These parameters not being well constrained by the analyzed datasets. For other densities such as $c + \bar c$, this parameter also determined with slightly large uncertainties showing that the heavy flavor tagged cross sections can not constrain these parameters well enough. However, we prefer to let these parameters to be free in the fit and in the FFs uncertainty determination to give more flexibility to the  parameterizations. 

In addition to the much more flexible input parametrization for {\it residual} charged hadrons FFs proposed in Sec.~\ref{sec:parametrization} (see Eq.~\eqref{parametrization}), we have repeated our QCD analysis with variety of alternative parameterizations, even more flexible than the one we finally used in our analysis.
For example, we have chosen the $(1 + \gamma_i z + \eta_i \sqrt{z})$, even allowing the fit to vary these new parameters. We also examine another parametrization form such as $D_{i}  (z,\mu_{0}^{2}) =N_{i} z^{\alpha_{i}}(1-z)^{\beta_{i}} [1-e^{-\gamma_{i} z}]$~\cite{Soleymaninia:2013cxa}. None of these modifications resulted in any significant improvement in $\chi^2$ optimizations, in the quality of the fit to the analyzed SIA data sets, or decreasing of the {\it residual} charged hadrons FFs uncertainty bands. This clearly indicates that the present SIA data sets are not really able to discriminate between various forms of the input distributions for the small {\it residual} charged hadrons FFs and the
stability of the corresponding FFs is not affected, as long as a sufficiently flexible choice is made. Therefore, we mainly focused on a very simple standard parameterizations for the {\it residual} charged hadrons FFs as presented in Eq.~\eqref{parametrization}. 

\begin{figure*}[htb]
\begin{center}
\vspace{0.50cm}
\resizebox{0.45\textwidth}{!}{\includegraphics{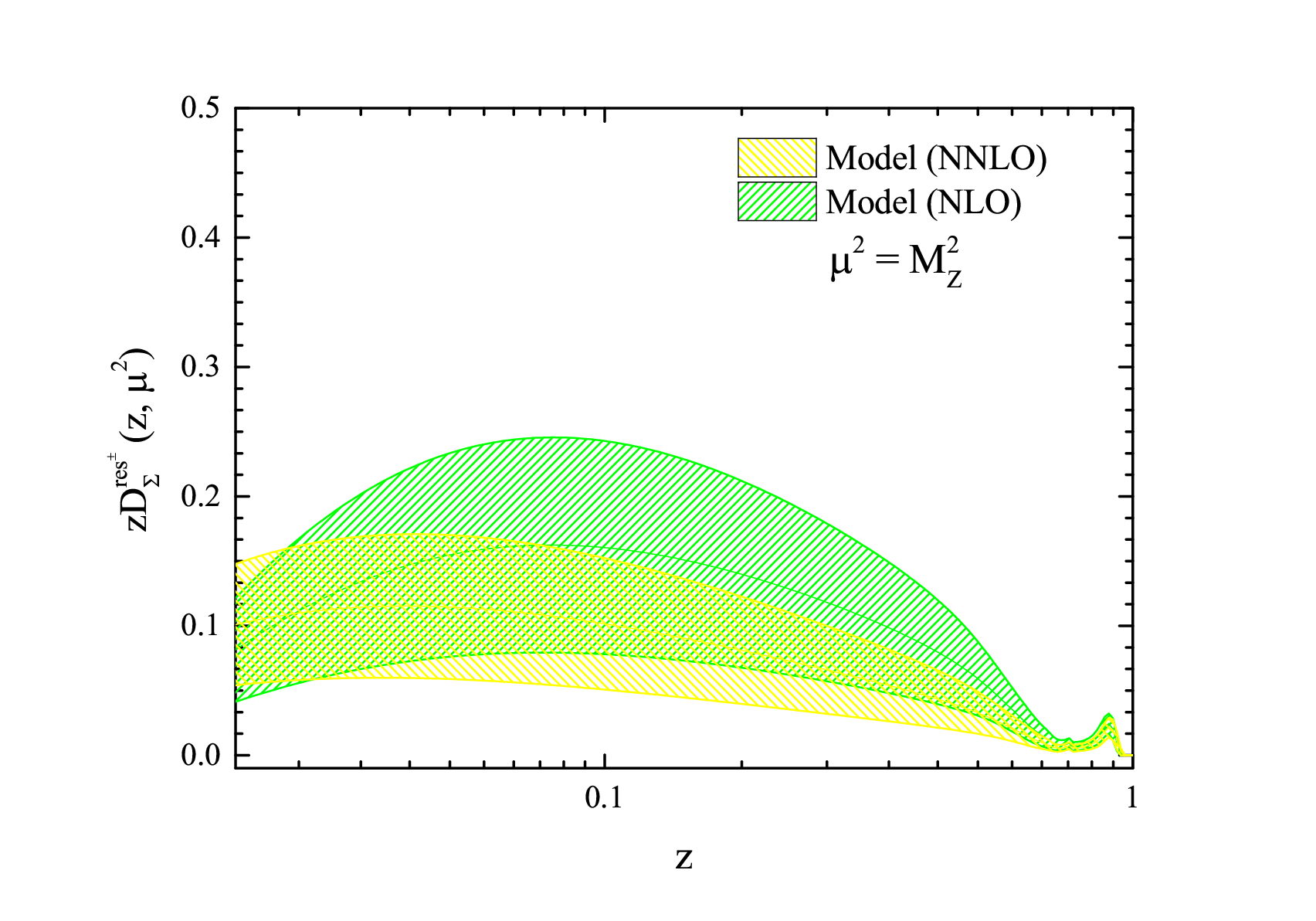}} 
\resizebox{0.45\textwidth}{!}{\includegraphics{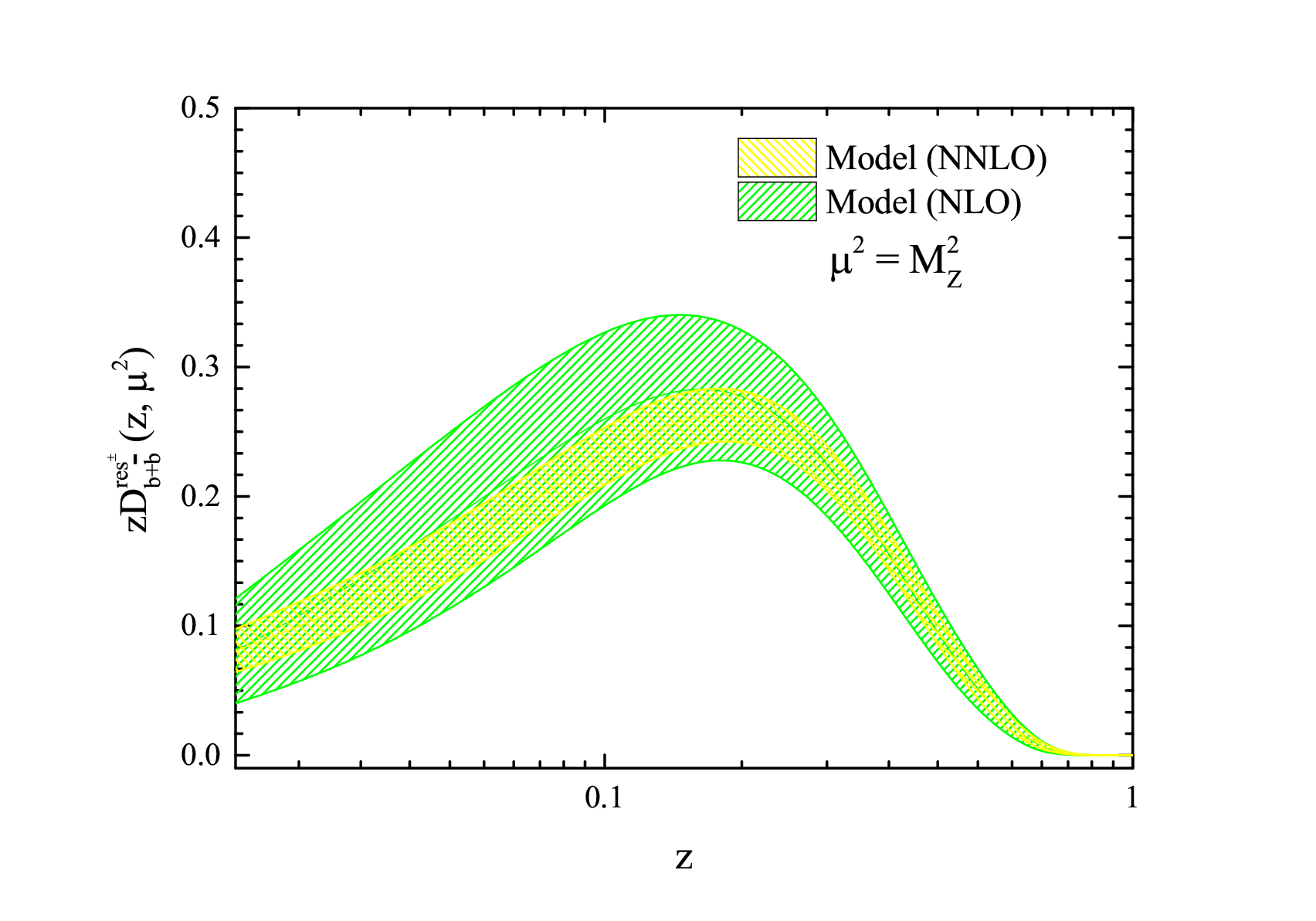}}  
\resizebox{0.45\textwidth}{!}{\includegraphics{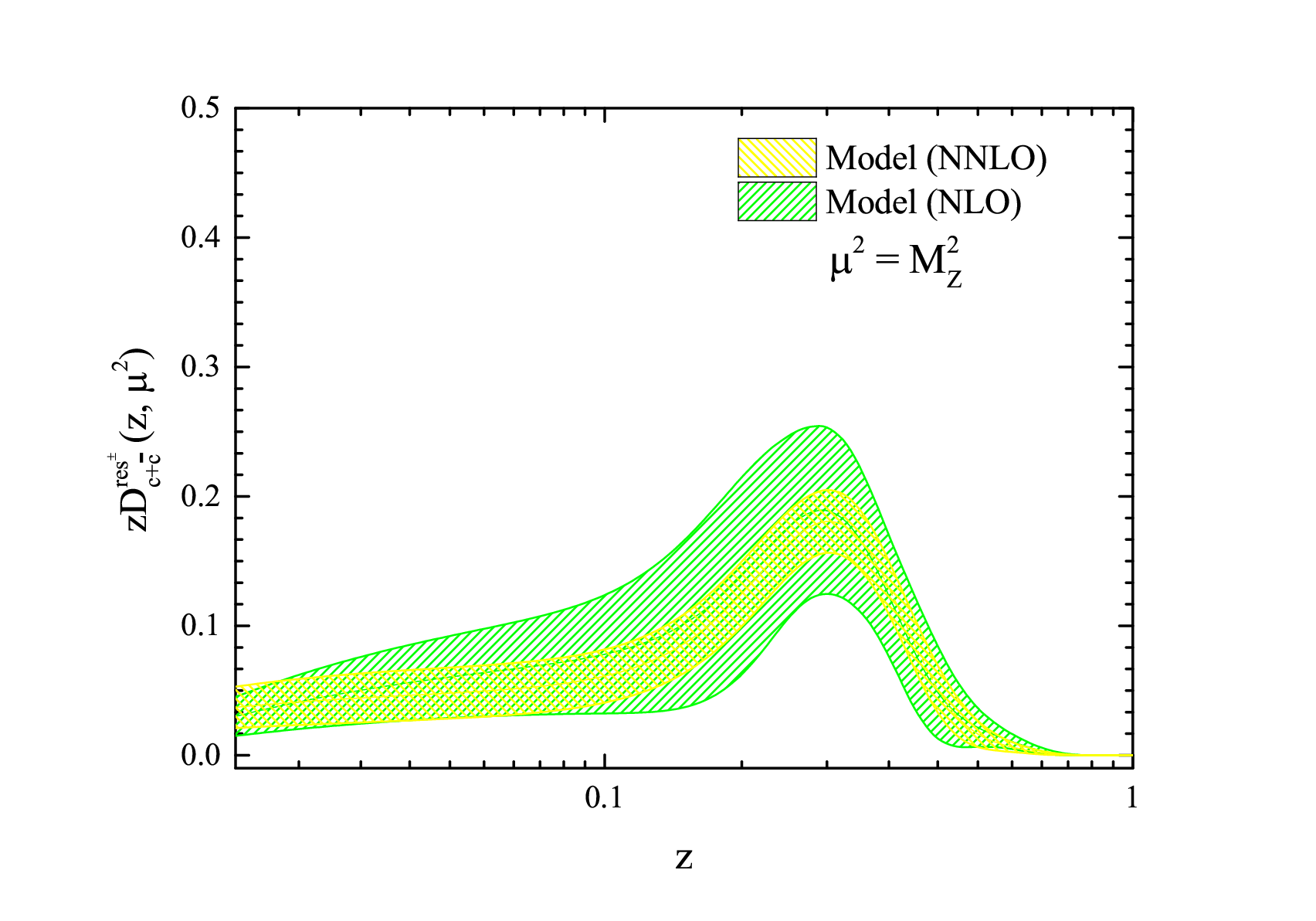}} 
\resizebox{0.45\textwidth}{!}{\includegraphics{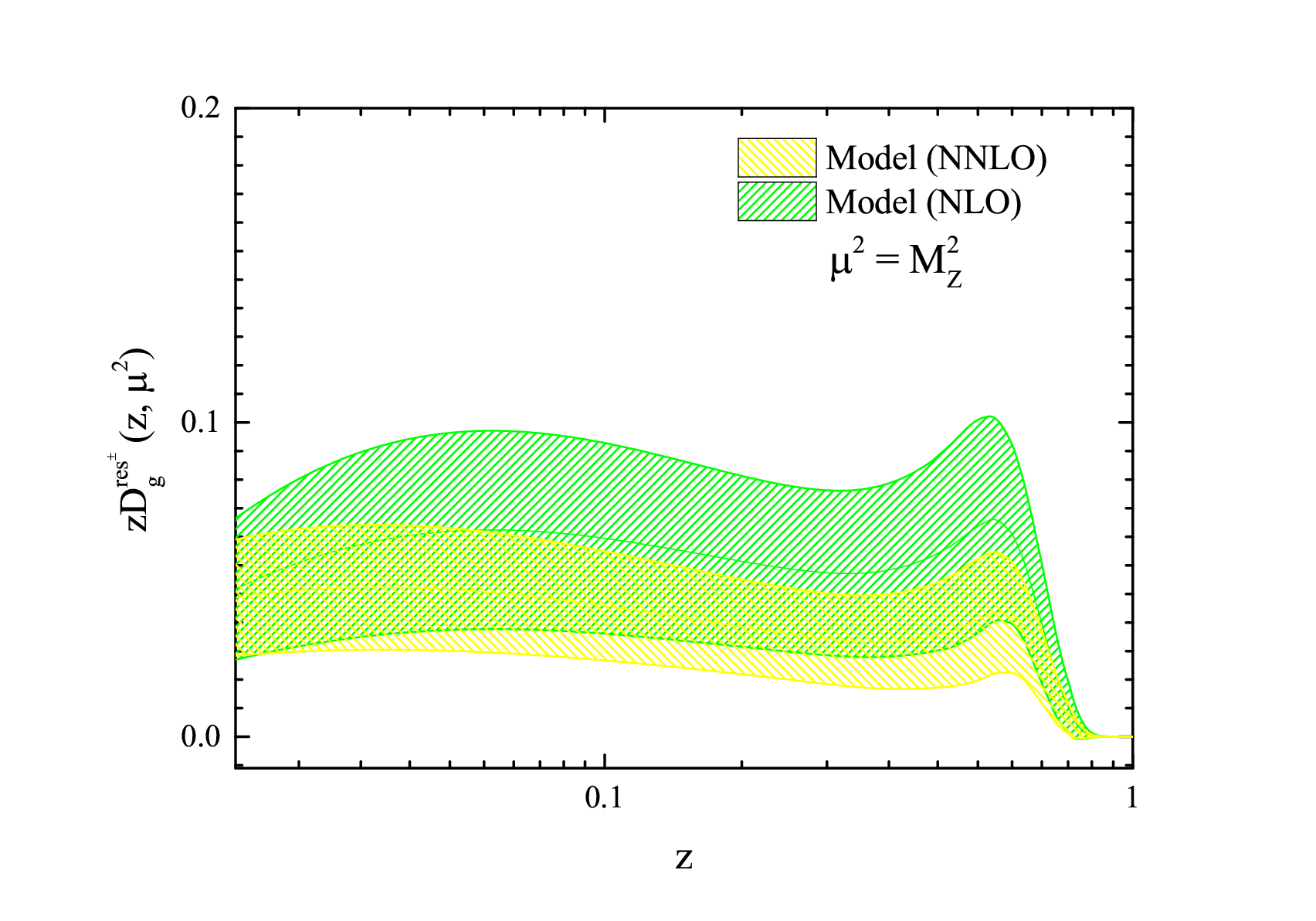}}
\caption{ {\it Residual } charged hadrons FFs determined from this analysis are shown for $zD^{\it {res}^\pm}_i (z, \mu^2)$ at the scale of $\mu^2 = M_Z^2$ for singlet, $c + \bar{c}$, $b + \bar{b}$, and gluon FFs at NLO and NNLO. The shaded bands provide uncertainty estimates using a criterion of $\Delta \chi^2 = 1$ as an allowed tolerance on the $\chi^2$ value of our QCD fit, as described in the text.  } \label{fig:FFs_Res_NNLO_error}
\end{center}
\end{figure*}

In Fig.~\ref{fig:FFs_Res_NNLO_error} we present the resulting {\it residual} charged hadrons FFs entitled ``{\tt Model}'' along with estimates of their uncertainty bands at NLO and NNLO accuracy. The resulting {\it residual} charged hadrons FFs $zD^{\it {res}^\pm}_i (z, \mu^2)$ are shown at the scale of $\mu^2 = M_Z^2$ for the singlet $\Sigma=u + \bar{u}+d + \bar{d}+s + \bar{s}$,  $c + \bar{c}$, $b + \bar{b}$, and gluon FFs at NLO and NNLO accuracy. The shaded bands provide uncertainty estimates using a criterion of $\Delta \chi^2 = 1$ as allowed tolerance on the $\chi^2$ value of our QCD fit. We have mentioned earlier that in our fit we consider the symmetric total up, down and strange distributions: $zD^{ {res}^\pm}_{u + \bar u}(z, \mu_0^2) = zD^{{res}^\pm}_{d + \bar d}(z, \mu_0^2) = zD^{{res}^\pm}_{s + \bar s}(z, \mu_0^2)=\frac{1}{3}zD^{{res}^\pm}_{\Sigma}(z, \mu_0^2)$.  
In this figure, the yellow bands represent the uncertainty for the NNLO accuracy and green bands indicate the uncertainty of NLO analysis. As one can see, considering the NNLO accuracy leads to a smaller FFs uncertainties. From Tables.~\ref{tab:NLO} and \ref{tab:NNLO} one also can conclude that the inclusion of higher order corrections leads to a smaller values of $\chi^2$.

%
\begin{table*}[t]
\caption{\label{tab:NLO} Fit parameters for the fragmentation of quarks and gluon into the $D^{{res}^\pm}$ at NLO accuracy.
The starting scale is taken to be $Q_{0}=5$~GeV for all parton species.
The values labeled by (*) have been fixed after the first minimization, since the available SIA data dose not constrain all unknown fit parameters well enough.  }
\begin{ruledtabular}
\begin{tabular}{cccccc}
flavor $i$        & ${N}_i$          &  $\alpha_i$          & $\beta_i$            \\    \hline
$u+ \overline{u}$ & $0.00131 \pm 0.0004$  &  $154.590^*$         & $13.684 \pm 1.602$   \\
$g$               & $0.0259 \pm 0.0119$   &  $27.363^*$          & $13.274 \pm 3.268$   \\
$c+ \overline{c}$ & $0.0474 \pm 0.0111$   &  $8.212 \pm 5.540$   & $20.941 \pm 12.163$  \\
$b+ \overline{b}$ & $0.103 \pm 0.0108$    &  $0.557 \pm 0.369$   & $5.365 \pm 0.730$    \\ 
$\chi^2/{\tt dof}$ &    $0.699$                                                      \\ 
$\alpha_s(M_Z^2)$   & $0.1176^*$~\cite{Bertone:2017tyb,Bertone:2018ecm}                 \\
$m_c$               & $1.51^*$~\cite{Bertone:2017tyb,Bertone:2018ecm}                   \\
$m_b$               & $4.92^*$~\cite{Bertone:2017tyb,Bertone:2018ecm}                   \\  
\end{tabular}
\end{ruledtabular}
\end{table*}
%
%

%
\begin{table*}[t]
\caption{\label{tab:NNLO} Same as Table~\ref{tab:NLO} but for the NNLO analysis.   }
\begin{ruledtabular}
\begin{tabular}{cccccc}
flavor $i$        & ${N}_i$          &  $\alpha_i$          & $\beta_i$             \\    \hline
$u+ \overline{u}$ & $0.00152 \pm 0.0003$  &  $153.479^*$         & $14.400 \pm 1.525$    \\
$g$               & $0.0453 \pm 0.0125$   &  $24.085^*$          & $12.124 \pm 2.205$    \\
$c+ \overline{c}$ & $0.0412 \pm 0.0108$   &  $9.680 \pm 7.503$   & $25.394 \pm 17.315$   \\
$b+ \overline{b}$ & $0.0977 \pm 0.0103$   &  $0.508 \pm 0.384$   & $5.751 \pm 0.855$     \\ 
$\chi^2/{\tt dof}$ &     $0.643$                                                      \\ 
$\alpha_s(M_Z^2)$   & $0.1176^*$~\cite{Bertone:2017tyb,Bertone:2018ecm}                  \\
$m_c$               & $1.51^*$~\cite{Bertone:2017tyb,Bertone:2018ecm}                    \\
$m_b$               & $4.92^*$~\cite{Bertone:2017tyb,Bertone:2018ecm}                    \\  
\end{tabular}
\end{ruledtabular}
\end{table*}
%
%

%
%
%
%

%
\subsection{ Charged hadrons FFs and comparison with other FF sets } 
%

\begin{figure*}[htb]
\begin{center}
\vspace{0.50cm}
\resizebox{0.45\textwidth}{!}{\includegraphics{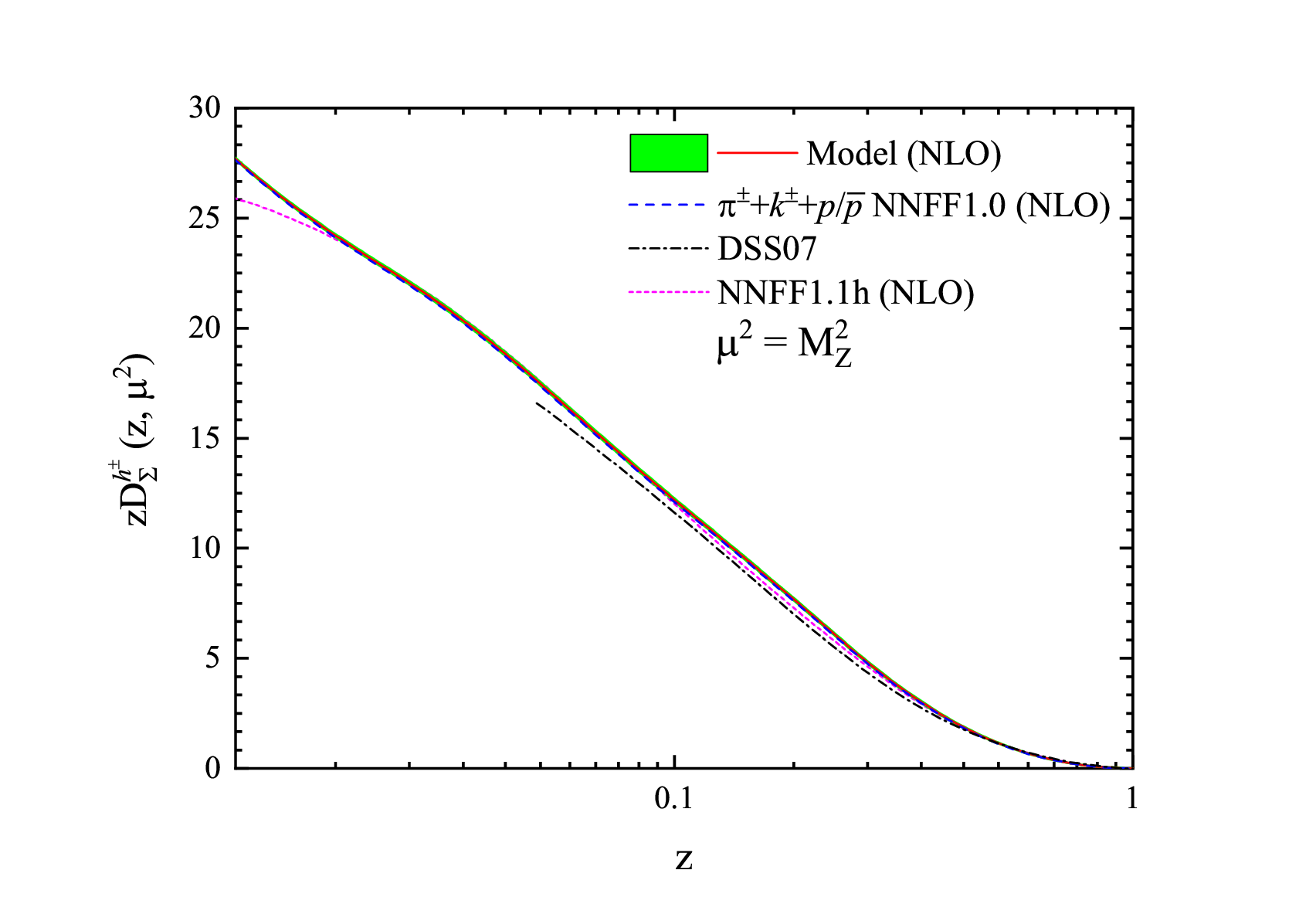}} 
\resizebox{0.45\textwidth}{!}{\includegraphics{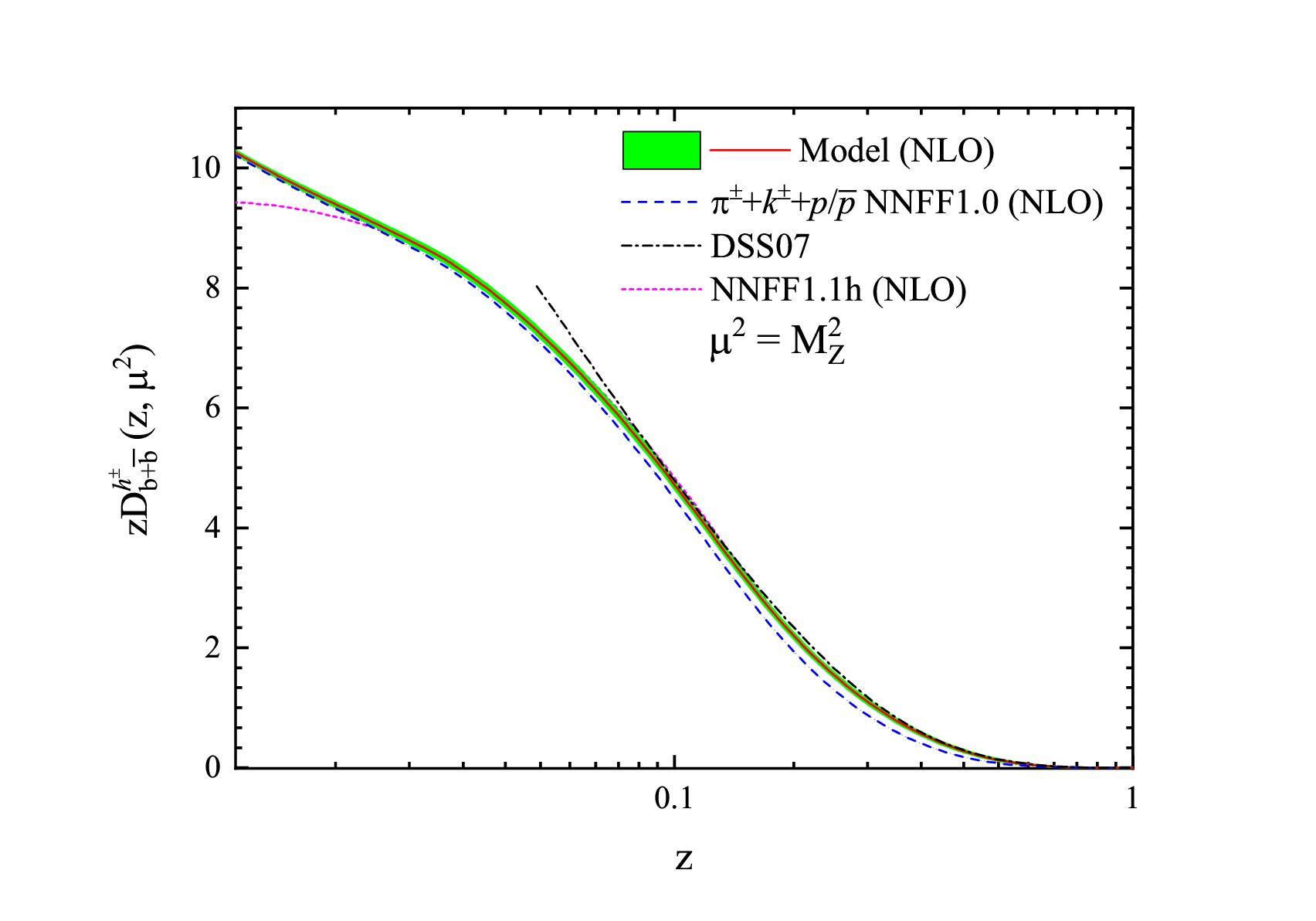}}  
\resizebox{0.45\textwidth}{!}{\includegraphics{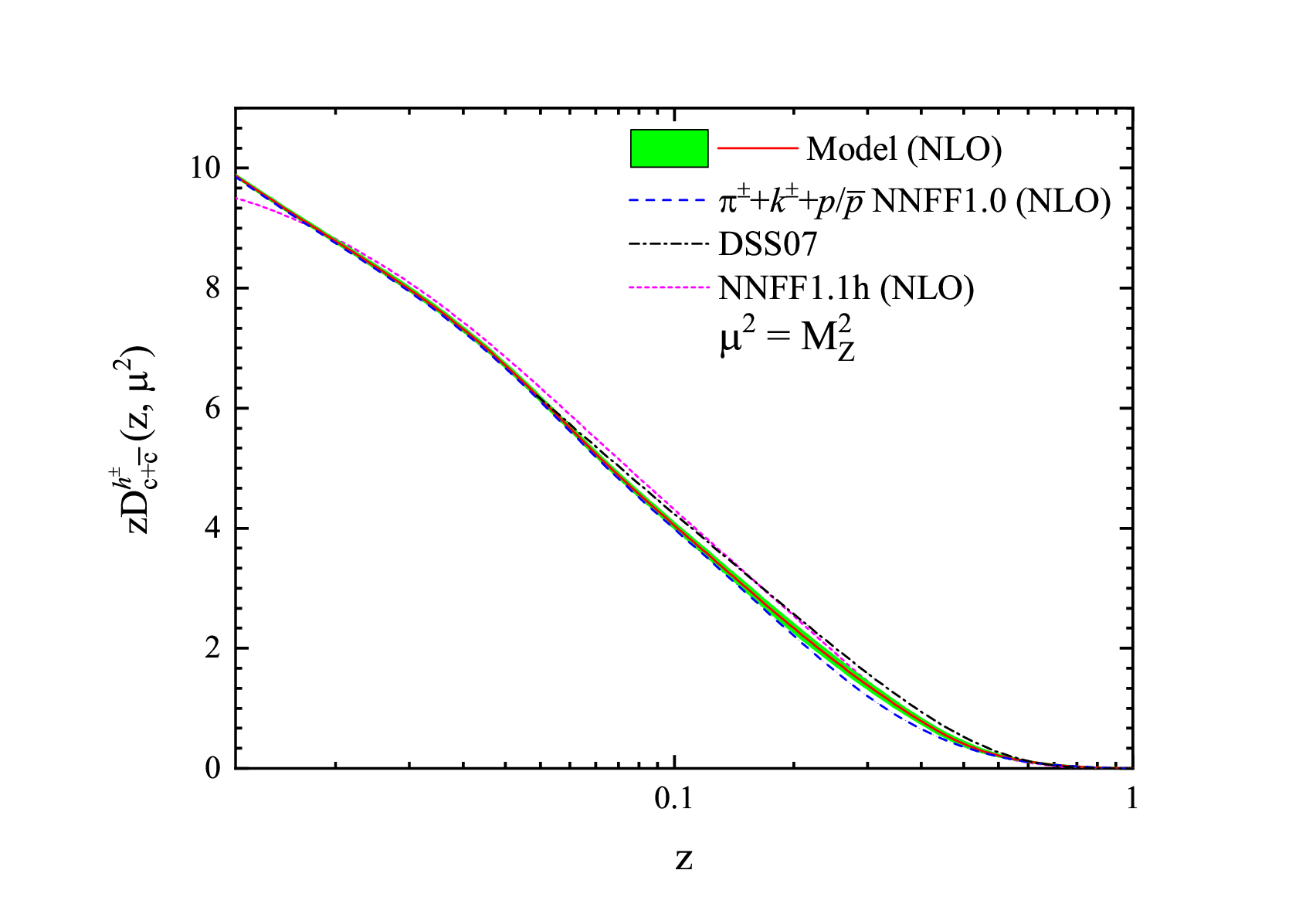}} 
\resizebox{0.45\textwidth}{!}{\includegraphics{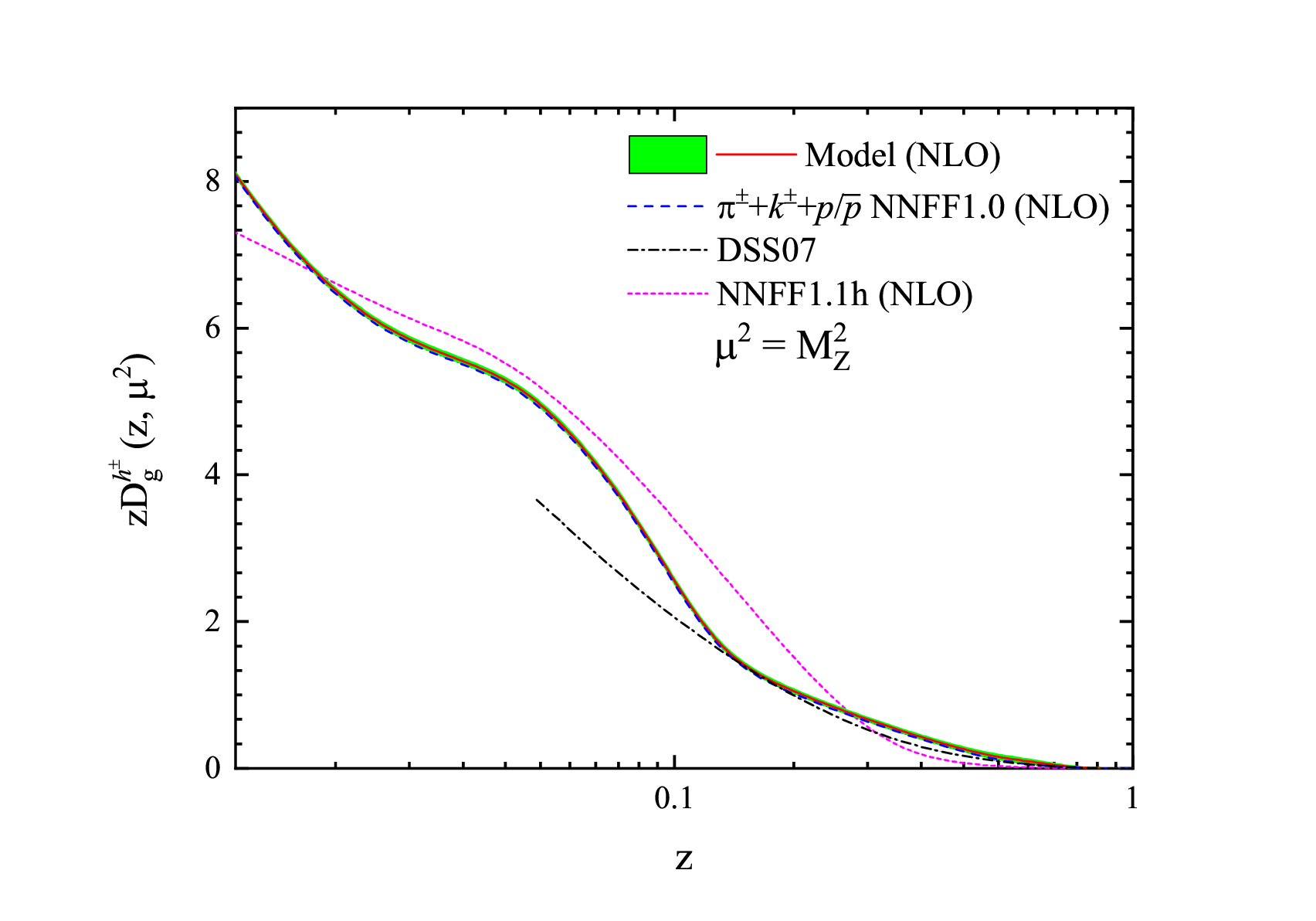}}
\caption{ Charged hadrons FFs determined from this analysis (solid line) are shown for $zD^{\it h^\pm}_i (z, \mu^2)$ at the scale of $\mu^2 = M_Z^2$ for $\Sigma$, $b + \bar{b}$, $c + \bar{c}$, and gluon FFs at NLO. The corresponding result from {\tt DSS07}~\cite{deFlorian:2007ekg} (dot-dashed lines) as well as the recent identified light FFs from {\tt NNFF1.0}~\cite{Bertone:2017tyb} (dashed lines) have been shown for comparison. Also our results are compared with the most recent unidentified charged hadron FFs from {\tt NNFF1.1h}~\cite{Bertone:2018ecm} (short-dashed lines). The shaded bands provide the uncertainty calculations of our {\it residual } charged hadrons FFs using a criterion of $\Delta \chi^2 = 1$ as an allowed tolerance on the $\chi^2$ value of our QCD fit.} \label{fig:FFs-MZ-NLO}
\end{center}
\end{figure*}

In this section, we give a detailed discussions of the first QCD analysis of {\it residual} charged hadrons FFs at NLO and NNLO which in the following will be referred to as ``{\tt Model}''. 
We now turn to present the charged hadrons FFs determined by using our {\it residual} charged hadrons  FFs in this analysis and compare it with other results in literature.
Firstly, in Fig.~\ref{fig:FFs-MZ-NLO} we compare our results for the unidentified charged hadron FFs at NLO (sum of our {\it residual} FFs with the $\pi^\pm, K^\pm ~ {\text{ and}} ~ p/\bar{p}$ FFs from {\tt NNFF1.0}) with those of the previous charged hadrons FFs, {\tt DSS07}~\cite{deFlorian:2007ekg}, as well as the most recent results from {\tt NNFF1.0} by NNPDF collaboration~\cite{Bertone:2017tyb}.
It should be mentioned here that the {\tt NNFF1.0} FFs are determined for the light identified charged hadrons of $\pi^\pm$, $K^\pm$ and $p/\bar{p}$ FFs. However the {\tt DSS07} FFs  for unidentified light charge hadrons are calculated by sum of the FFs from {\it residual} and $\pi^\pm, K^\pm ~ {\text{ and}} ~ p/\bar{p}$. 
In order to present the impact of our {\it residual} FFs  in calculation of unidentified charged hadron FFs, we calculate the total {\tt NNFF1.0} FFs for charged pion, kaon and (anti) proton  entitled  as ``{\tt $\pi^\pm + K^\pm + p/\bar{p}$ NNFF1.0}" and compare with other FF sets.

Since in our analysis we parameterize the $q+\bar{q}$ combinations for FFs, and hence, the $u+\bar{u}, d+\bar{d}, s+\bar{s}, c+\bar{c}, b+\bar{b}$ and $g$ FFs can be compared directly to those of other analyses in literature. The comparison in Fig.~\ref{fig:FFs-MZ-NLO} is shown at $Q^2 = M_Z^2$ for the NLO analysis. We should mentioned here that, in order to quantitatively assess the impact of the contribution from light quark and antiquark FFs for the {\it residual} charged hadrons, in Fig.~\ref{fig:FFs-MZ-NLO} we plot the total light quarks and antiquarks contributions $D^{h^\pm}_\Sigma$.

The main differences between our charged hadrons FFs results and {\tt DSS07} are found for the gluon FF in the region $z<0.1$. In this region, the gluon FFs from {\tt DSS07} is smaller than our gluon FFs. At small to large values of $z$, in the kinematical coverage of the SIA data sets, as expected, our charged hadrons FFs and {\tt $\pi^\pm+k^\pm+p/\bar{p}$ NNFF1.0} for $D^{h^\pm}_\Sigma$ and gluon are statistically equivalent.

On the other hand, our total heavy quark-antiquark combinations $c+\bar{c}$ and  $b+\bar{b}$ FFs are only moderately affected by the {\it residual} contributions, which leads to a minor enhancement in  comparison with {\tt $\pi^\pm+k^\pm+p/\bar{p}$ NNFF1.0} mostly for the whole $z$ region. For these distributions, the inclusion of {\it residual} contributions visibly affects the shape of this distribution  and the small contributions from {\it residual} charged hadrons FFs are evident.

It should be mentioned here that the {\tt NNFF1.1h} FFs for unidentified charged hadrons have been determined independently from {\it residual} and other light ($\pi^\pm, k^\pm ~ {\text{ and}} ~ p/\bar{p}$) FFs~\cite{Bertone:2018ecm}.  The comparison of our results with {\tt NNFF1.1h}, Fig.~\ref{fig:FFs-MZ-NLO}, shows that there are no big difference for the $b+\bar{b}$ and $D^{h^\pm}_\Sigma$ FFs, except for very small values of $z$. For the $c+\bar{c}$ FF, we see a small enhancement for the {\tt NNFF1.1h} for all range of momentum fraction $z$. A relatively big difference has been found for the gluon FF, specially for the region of $z < 0.3$.

\begin{figure*}[htb]
\begin{center}
\vspace{0.50cm}
\resizebox{0.45\textwidth}{!}{\includegraphics{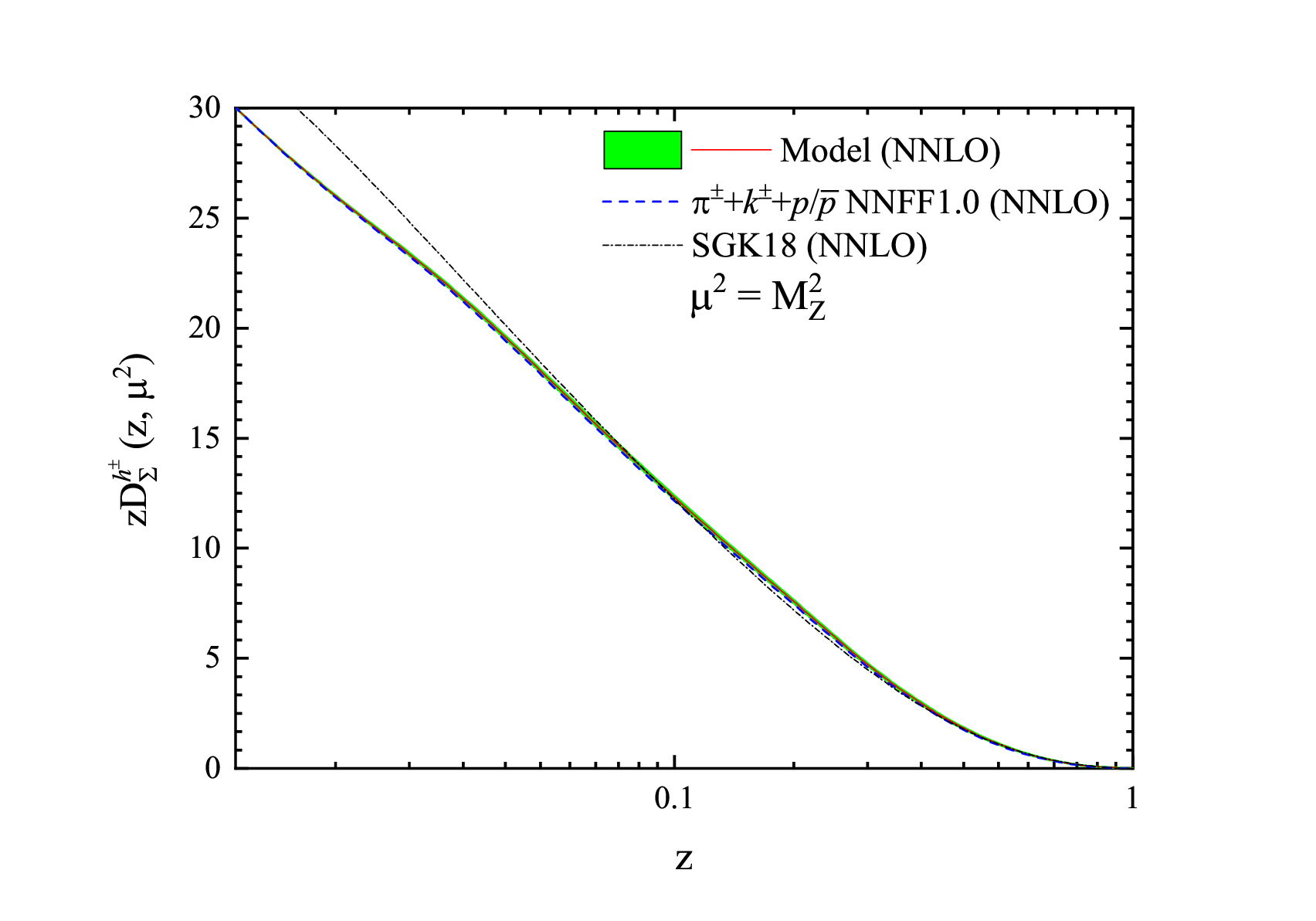}}  
\resizebox{0.45\textwidth}{!}{\includegraphics{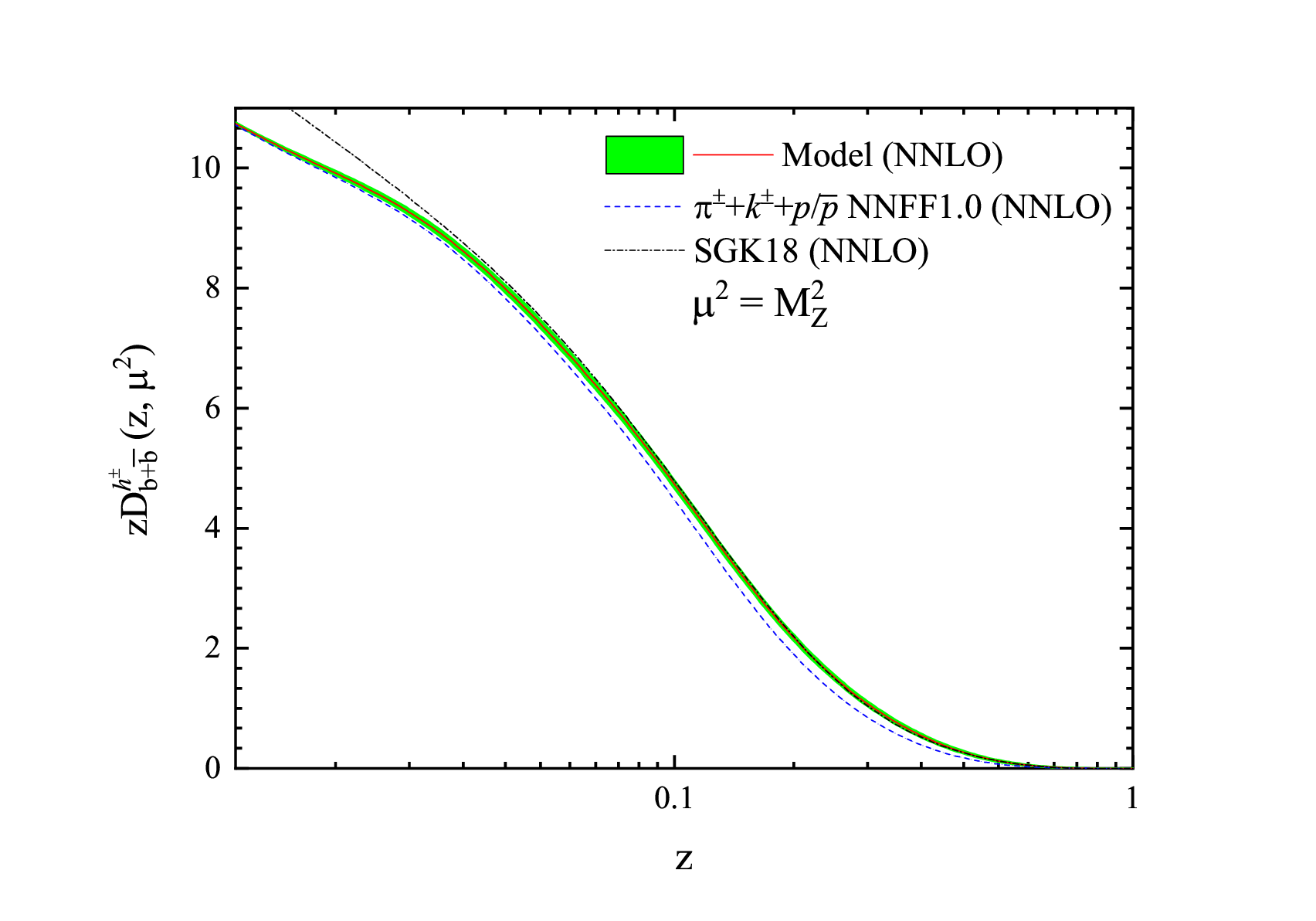}} 
\resizebox{0.45\textwidth}{!}{\includegraphics{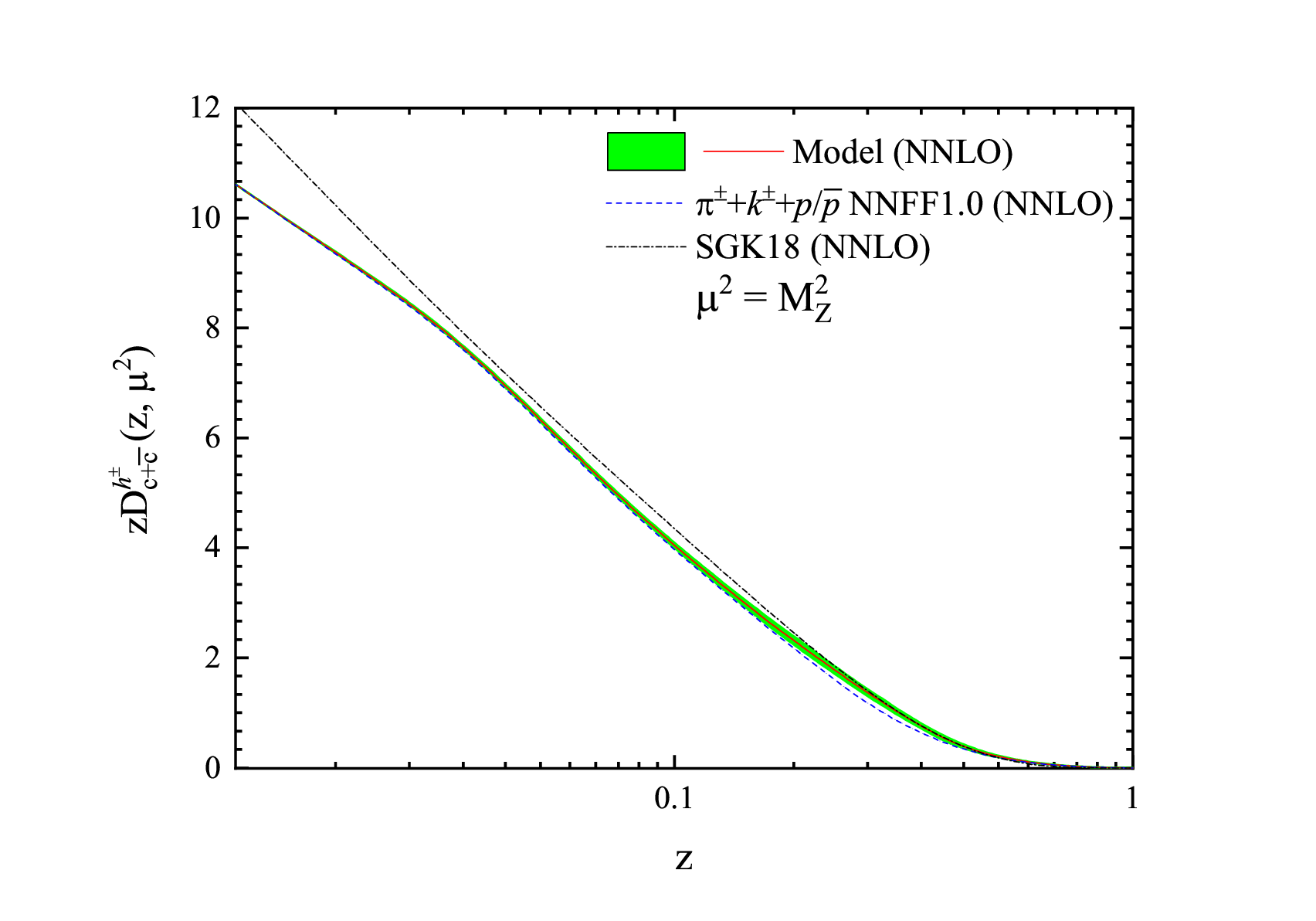}}  
\resizebox{0.45\textwidth}{!}{\includegraphics{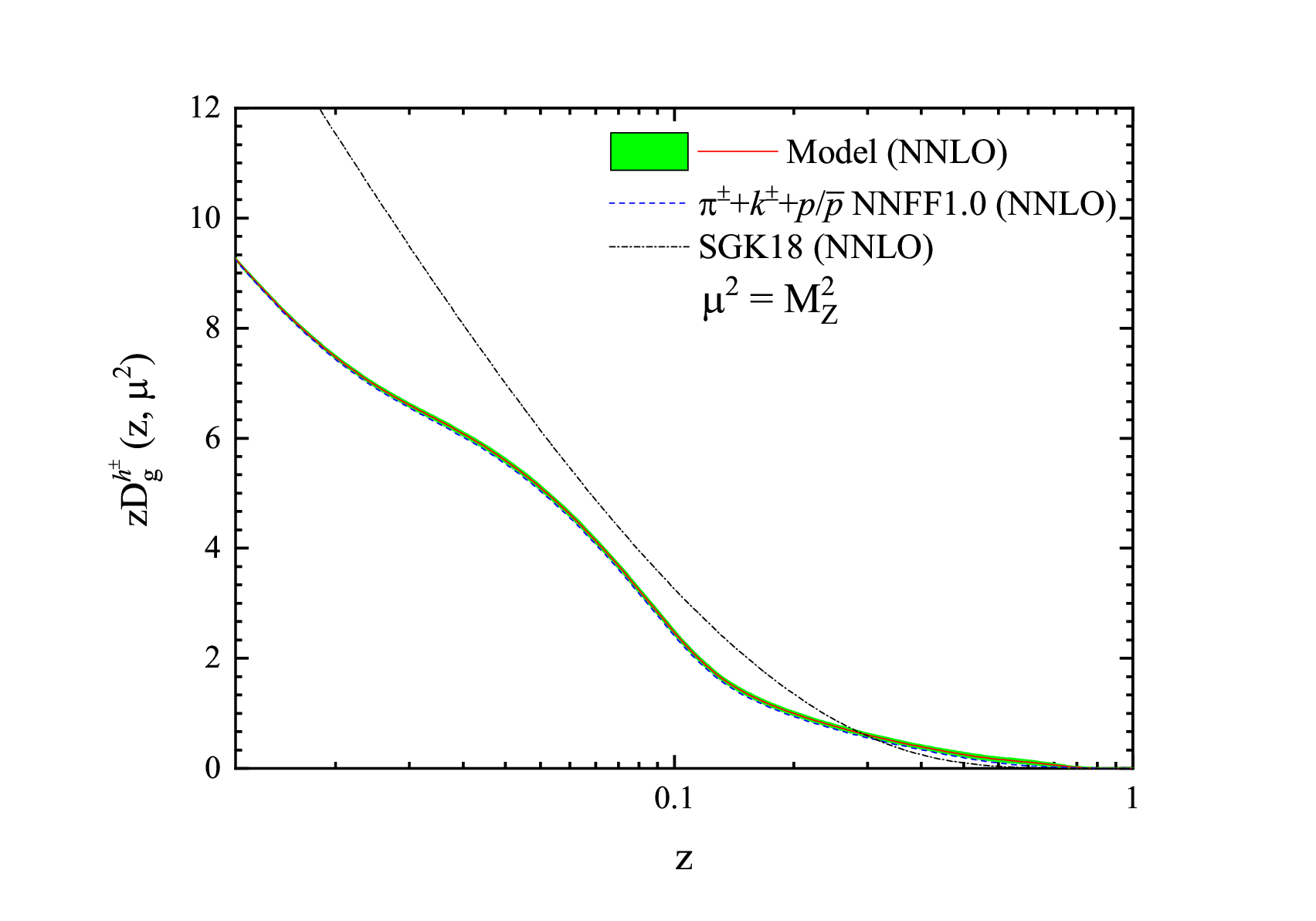}} 
\caption{ Charged hadrons FFs determined from this analysis (solid line) are shown for $zD^{\it h^\pm}_i (z, \mu^2)$ at the scale of $\mu^2 = M_Z^2$ for $\Sigma$, $b + \bar{b}$, $c + \bar{c}$, and gluon FFs at NNLO accuracy. The corresponding result from the most recent {\tt SGK18}~\cite{Soleymaninia:2018uiv} (dot-dashed lines) as well as the recent identified light FFs from {\tt NNFF1.0}~\cite{Bertone:2017tyb} (dashed lines) have been shown for comparison. The shaded bands provide the uncertainty calculations of our {\it residual } charged hadrons FFs using a criterion of $\Delta \chi^2 = 1$ as an allowed tolerance on the $\chi^2$ value of our QCD fit.} \label{fig:FFs-MZ-NNLO}
\end{center}
\end{figure*}

We now turn to the NNLO charged hadrons FFs, which are compared in Fig.~\ref{fig:FFs-MZ-NNLO} to those from the $\pi^\pm+k^\pm+p/\bar{p}$ {\tt NNFF1.0} SIA QCD fit as well as the most recent analysis of {\tt SGK18}~\cite{Soleymaninia:2018uiv} at $Q^2 = M_Z^2$. The main conclusions from the comparison in Fig.~\ref{fig:FFs-MZ-NNLO} are the following: 
The $D^{h^\pm}_\Sigma$ and gluon FFs of our analysis at NNLO accuracy and {\tt $\pi^\pm+k^\pm+p/\bar{p}$ NNFF1.0} are qualitatively similar, which indicate a very small contributions from the {\it residual} charged hadrons FFs. Note however that for the case of heavy flavors, $c+\bar{c}$ and $b+\bar{b}$ FFs the difference are a little bigger, especially for the case of bottom quark FF over the whole range of $z$.
As already noticed in the discussion of our NLO results, inclusion of the {\it residual} charged hadrons FFs in a FFs analysis visibly affects the shape of the  $c+\bar{c}$ and $b+\bar{b}$ distributions. The comparison of our results with those of {\tt SGK18} are also shown in Fig.~\ref{fig:FFs-MZ-NNLO}. As one can see, all the distributions obtained by {\tt SGK18} are larger than our results for all range of $z$. Big difference for the gluon FF  between these two analyses are also evident from Fig.~\ref{fig:FFs-MZ-NNLO}. In view of the comparison with other charged hadrons FF sets and clear evidence of different shape of the heavy flavor distributions, it is interesting to consider the small contributions of {\it residual} charged hadrons FFs in any global QCD analysis of FFs.

As a final point, we should mentioned here that the uncertainty bands presented in Figs.~\ref{fig:FFs-MZ-NLO} and \ref{fig:FFs-MZ-NNLO} are only correspond to the uncertainty calculations of our {\it residual } charged hadrons FFs as indicated in Fig.~\ref{fig:FFs_Res_NNLO_error}.

%
%
%
%

%
%
%
%

In the next section, we compute the theory predictions for the SIA processes based on our results for the charged hadrons FFs, and compare results to the analyzed data sets. In addition, in order to discuss the size of contributions from {\it residual } charged hadrons FFs, we also present the data/theory ratio based on the extracted {\it residual } charged hadrons FFs.

%
%
\begin{figure*}[htb]
	\begin{center}
		\vspace{0.50cm}
		\resizebox{0.95\textwidth}{!}{\includegraphics{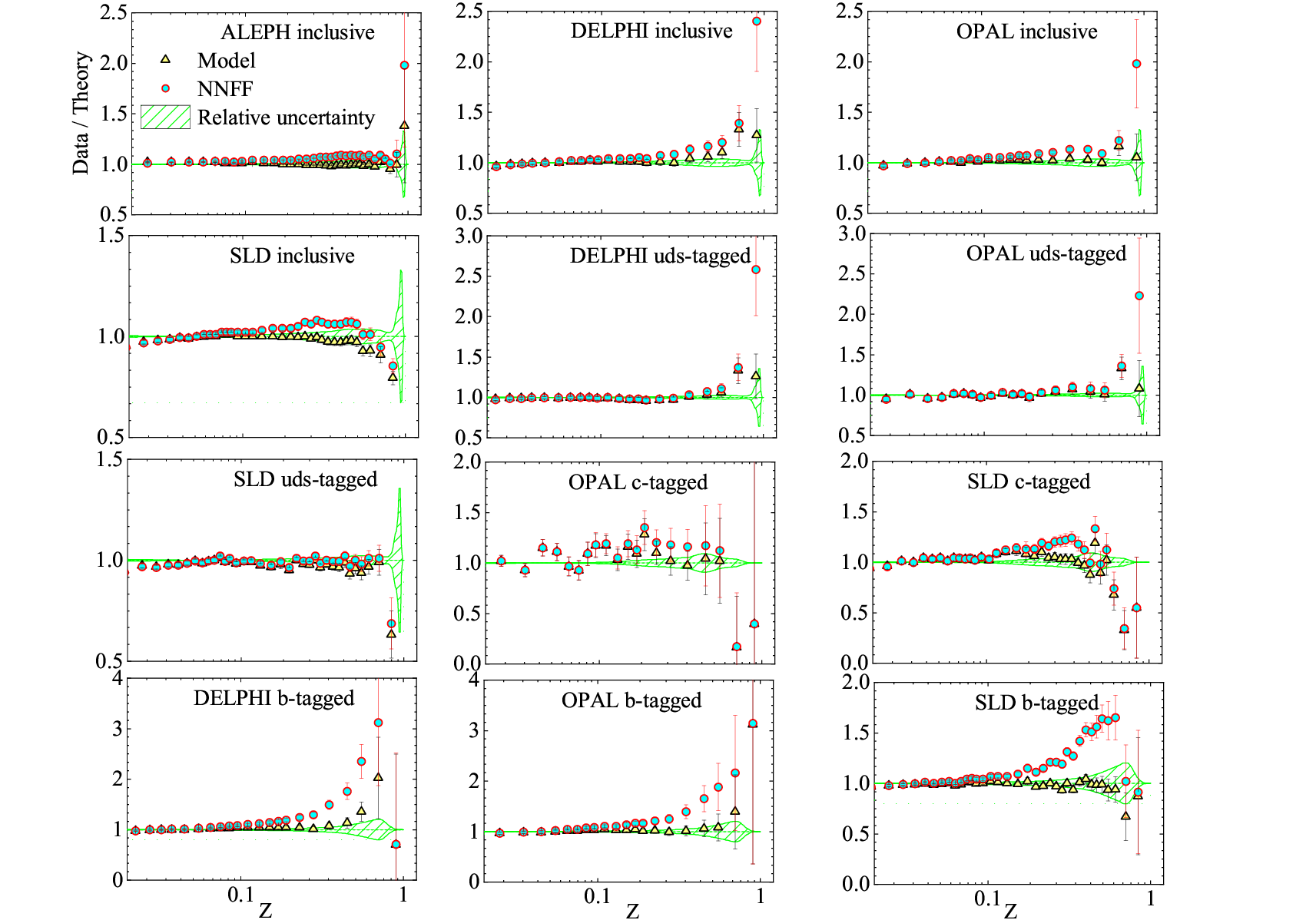}}  
		\caption{ The data/theory for our NNLO results for each of the data sets used in this analysis. The ratios are presented for the total inclusive, light, heavy quark $c$- and $b$-tagged normalized cross sections. The uncertainty bands originate from the uncertainty calculations of our {\it residual } charged hadrons FFs.  } \label{fig:data_theory}
	\end{center}
\end{figure*}
%
%

%
%
\begin{figure*}[htb]
	\begin{center}
		\vspace{0.50cm}
		\resizebox{0.95\textwidth}{!}{\includegraphics{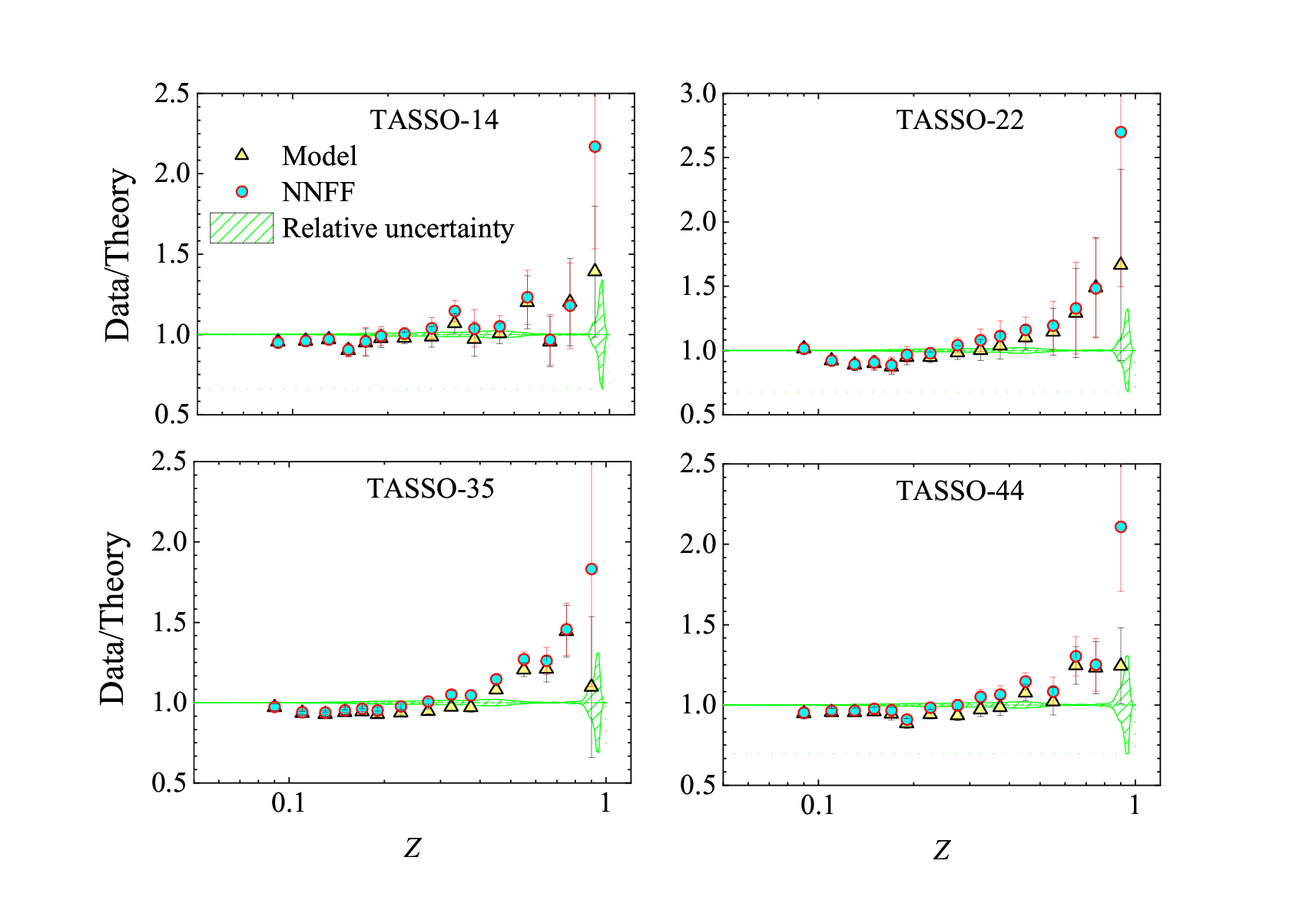}}  
		\caption{  Same as Fig.~\ref{fig:data_theory} but this time for the {\tt TASSO } data sets with the smaller values of center-of-mass energy, $\sqrt{s} = 14\,, 22\,, 35$ and $44$ GeV.  } \label{fig:data_theory_2}
	\end{center}
\end{figure*}
%
%

%
%
\begin{figure*}[htb]
	\begin{center}
		\vspace{0.50cm}
		\resizebox{0.45\textwidth}{!}{\includegraphics{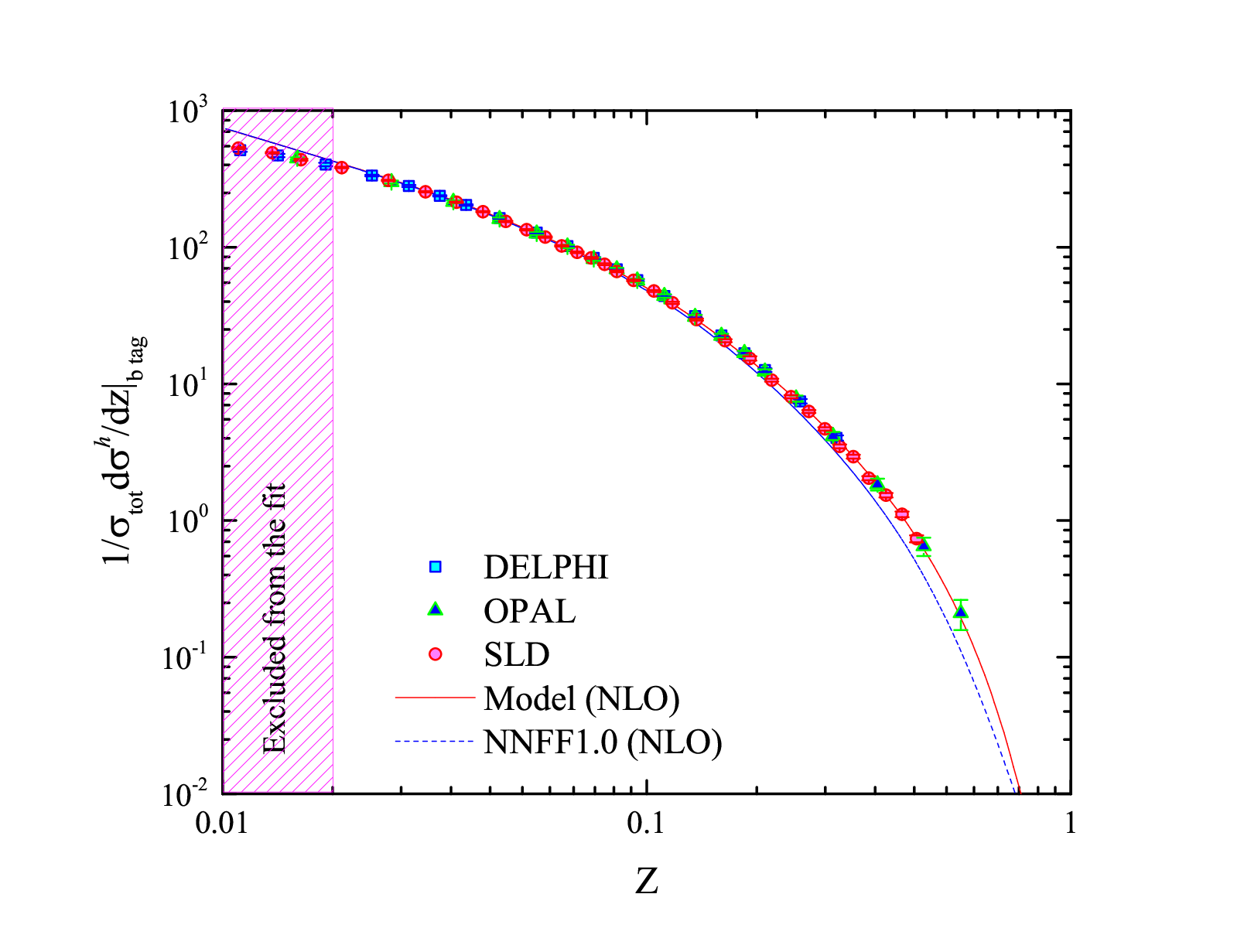}}  
		\resizebox{0.45\textwidth}{!}{\includegraphics{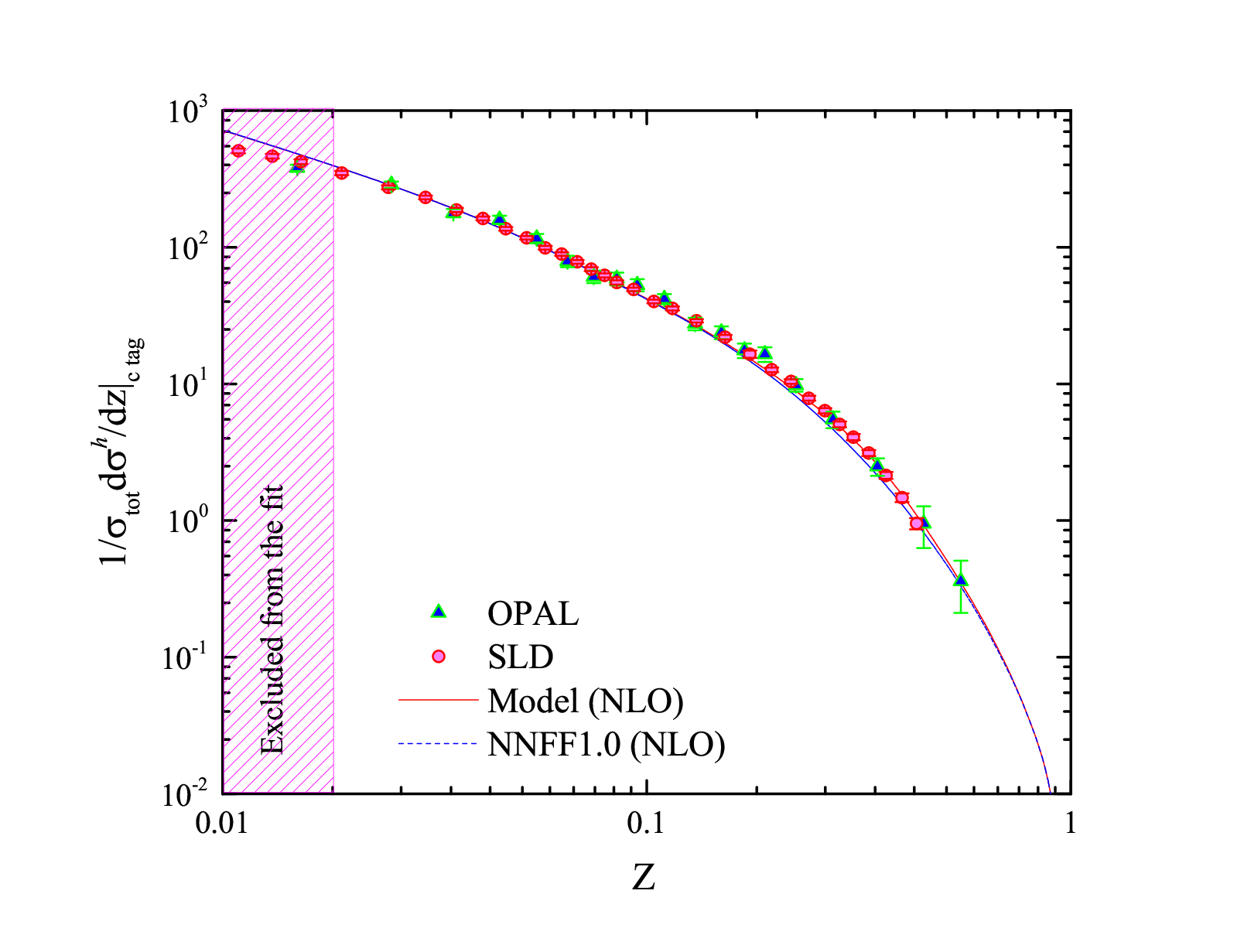}} 
		\resizebox{0.45\textwidth}{!}{\includegraphics{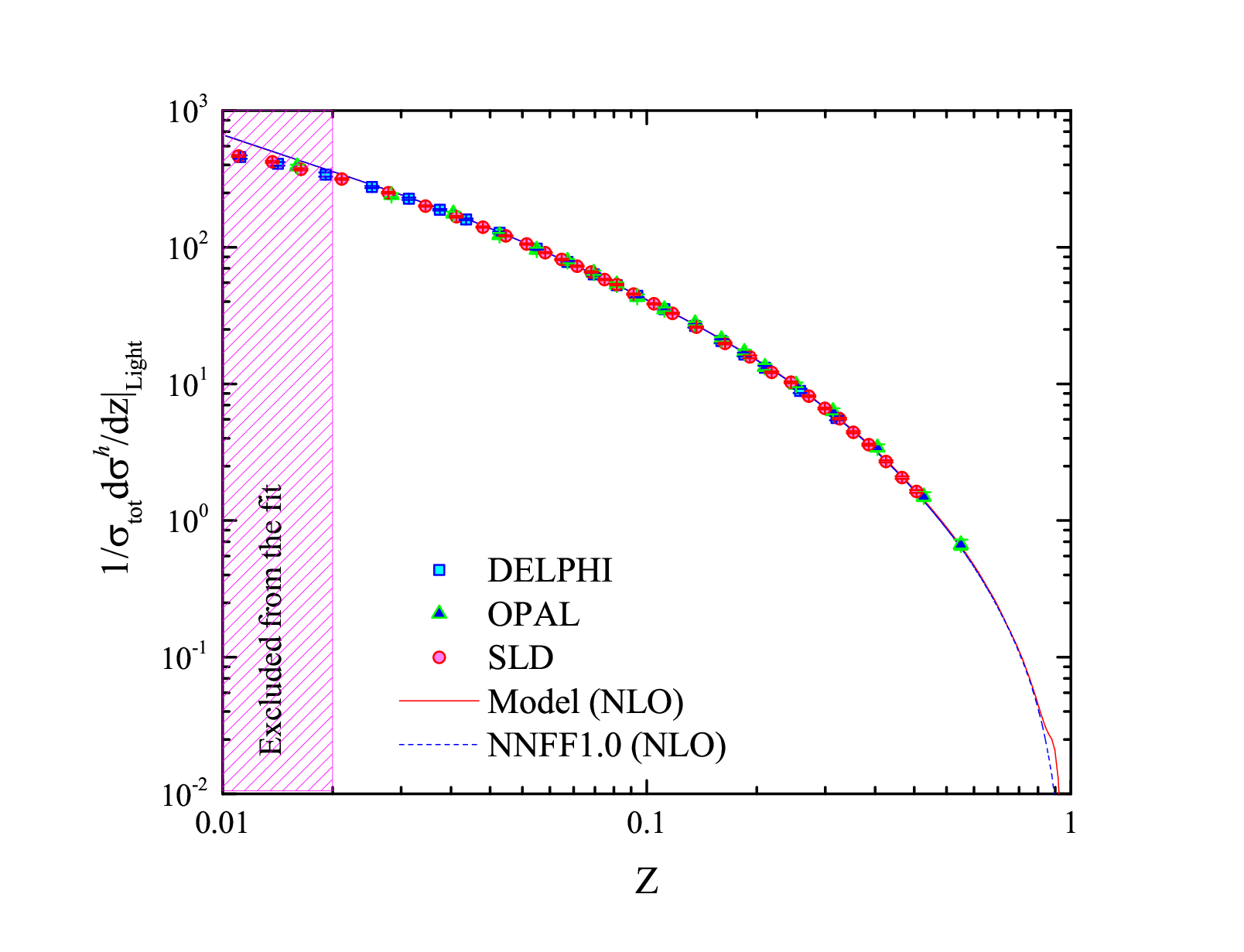}}  
		\resizebox{0.45\textwidth}{!}{\includegraphics{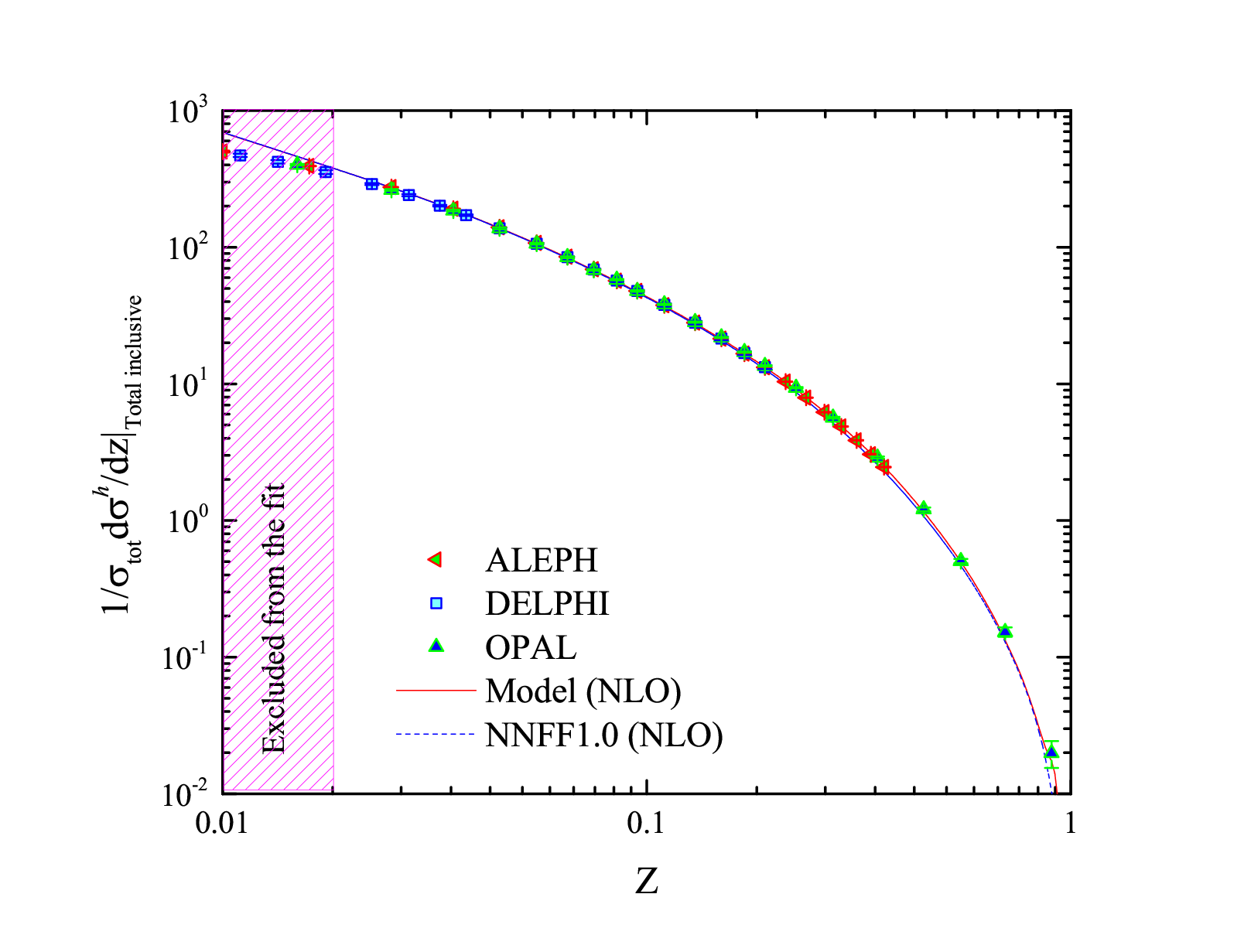}} 
		\caption{ SIA data sets compared to the best-fit results of our NLO QCD analysis of {\it residual } charged hadrons FFs ({\tt Model}; solid lines) for variety of SIA observables including total inclusive, $c$-tagged, $b$-tagged normalized cross sections. The theory predictions ({\tt NNFF1.0}; dashed lines) based on very recent {\tt NNPDF} collaboration~\cite{Bertone:2017tyb} have also been shown as well. } \label{fig:data-theory-NLO}
	\end{center}
\end{figure*}
%
%

%
%
\begin{figure*}[htb]
	\begin{center}
		\vspace{0.50cm}
		\resizebox{0.45\textwidth}{!}{\includegraphics{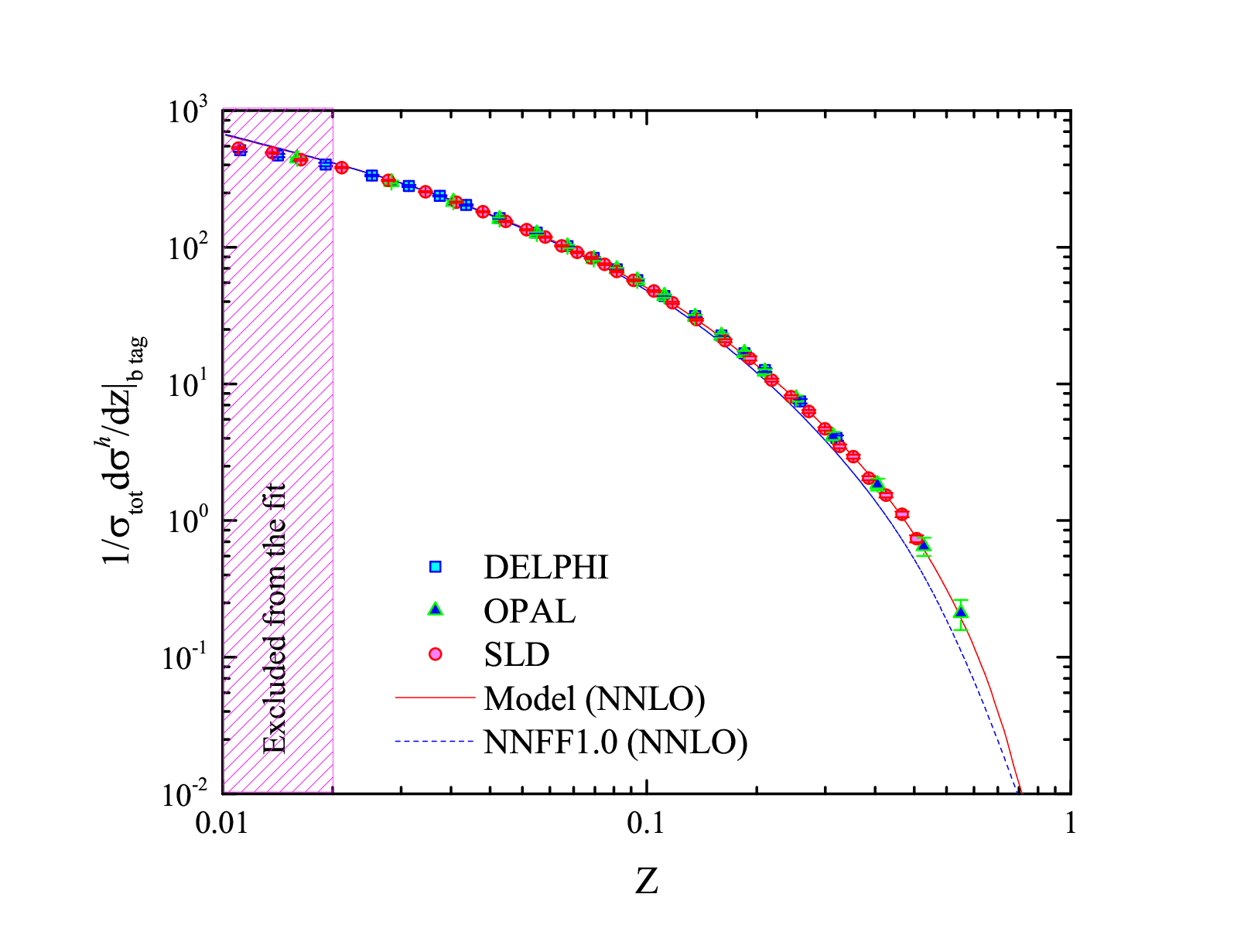}}  
		\resizebox{0.45\textwidth}{!}{\includegraphics{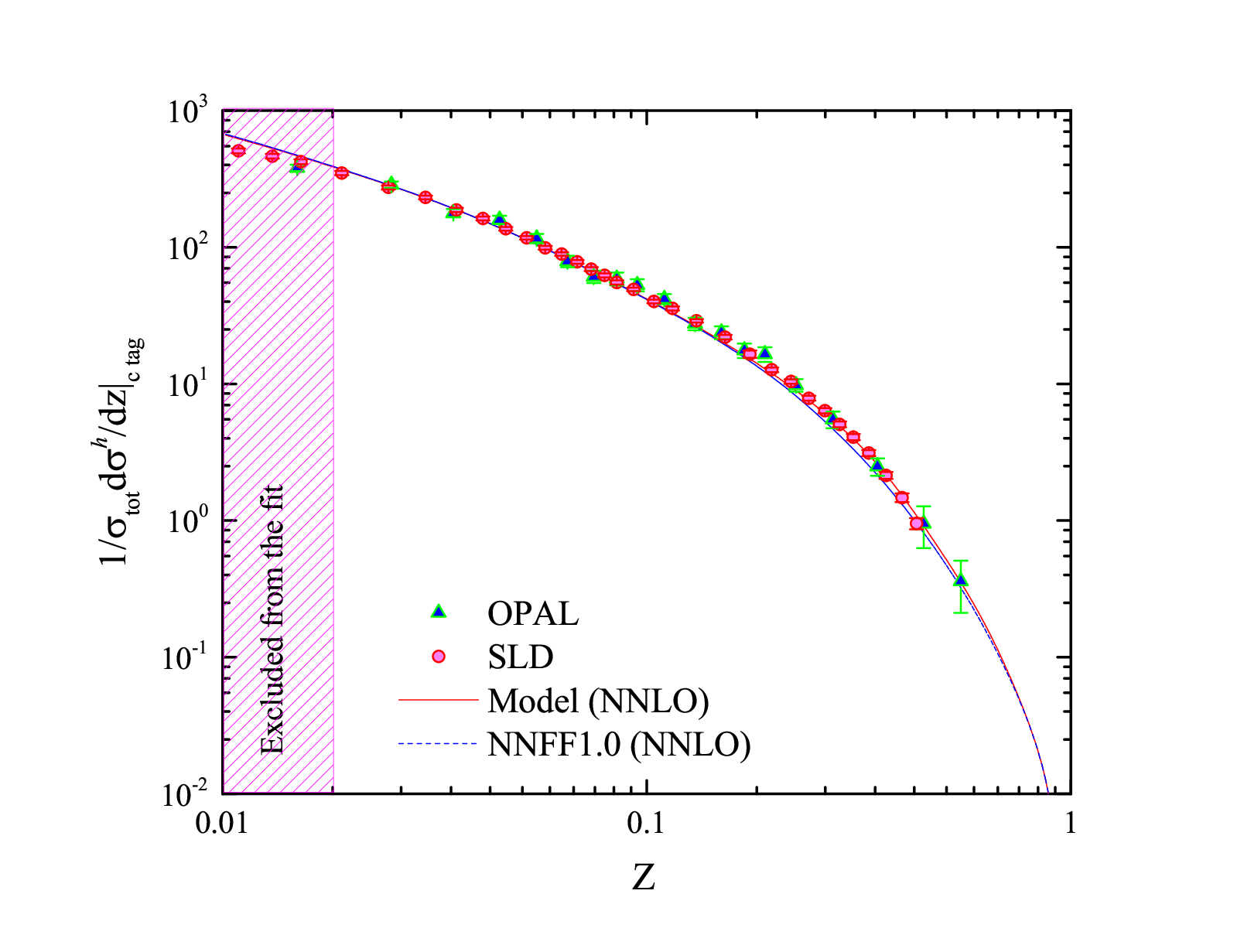}} 
		\resizebox{0.45\textwidth}{!}{\includegraphics{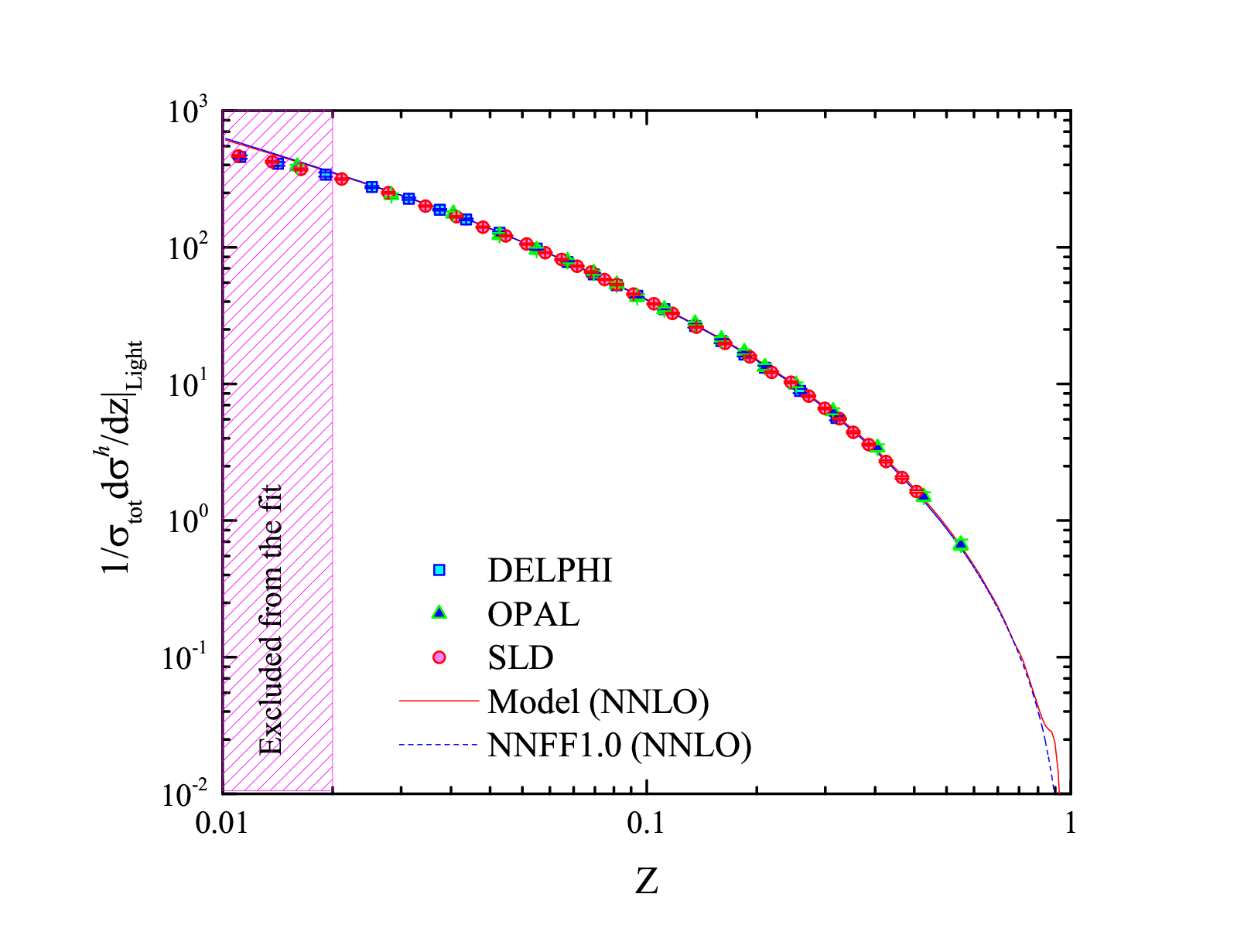}} 
		\resizebox{0.45\textwidth}{!}{\includegraphics{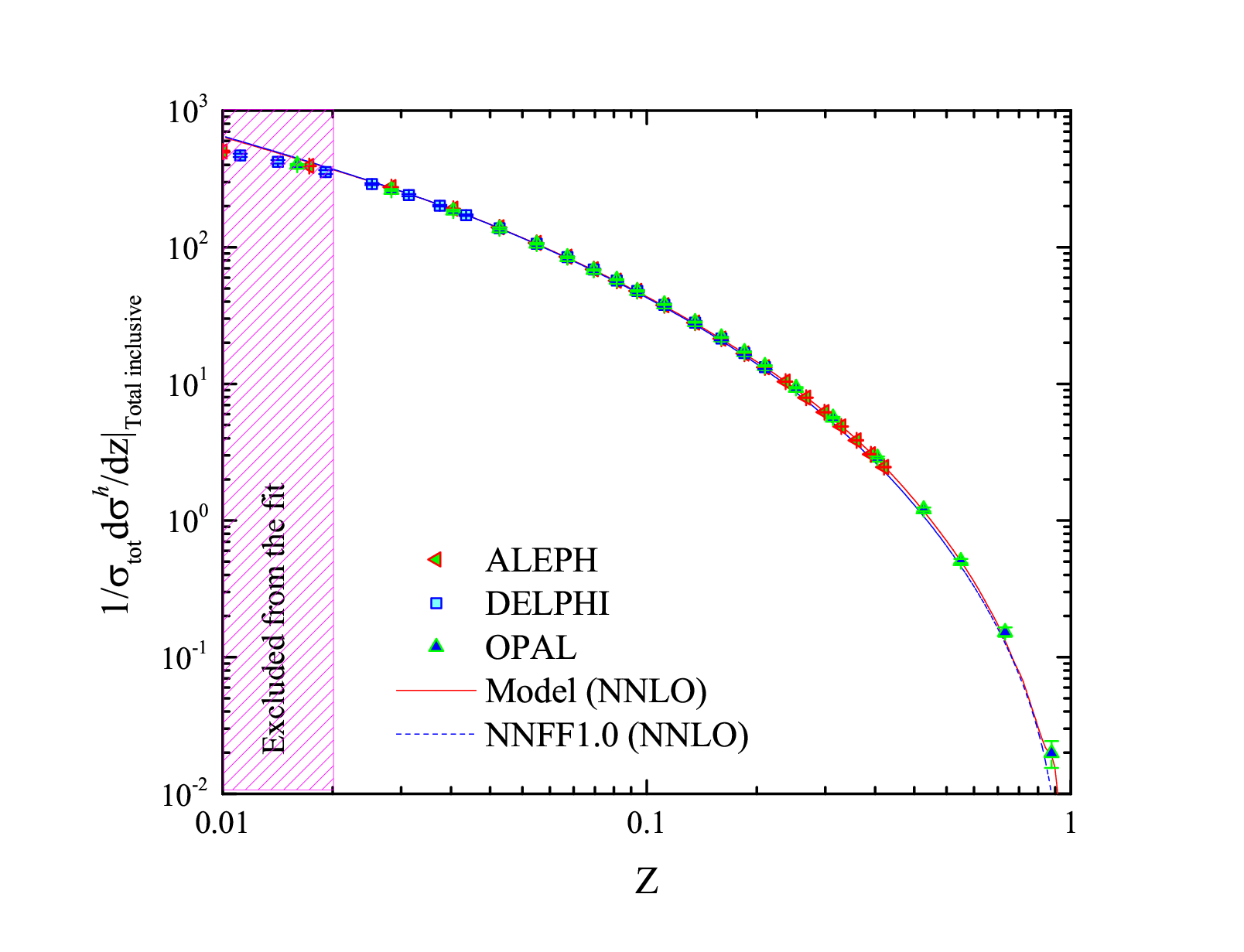}} 
		\caption{ Same as Fig.~\ref{fig:data-theory-NLO} but for NNLO accuracy. } \label{fig:data-theory-NNLO}
	\end{center}
\end{figure*}
%
%

%
\subsection{ Discussion of fit quality and data/theory comparison } 
%

After our detailed discussion on the determined {\it residual} charged hadrons FFs and details presentation as well as comparison with other results in literature, we are now in position to present our theory prediction using the extracted charged hadrons FFs.
In order to discuss the size of contributions from {\it residual } charged hadrons FFs, we present in Fig.~\ref{fig:data_theory} the data/theory ratio based on the extracted {\it residual } charged hadrons FFs at NNLO accuracy. These ratios are presented for the total inclusive, light, heavy quark $c$- and $b$-tagged normalized cross sections  at $\sqrt{s} = 91.2$ GeV. In Fig.~\ref{fig:data_theory_2}, we also present the data/theory for our results at NNLO accuracy for the {\tt TASSO } data sets which correspond to the smaller values of center-of-mass energy, $\sqrt{s} = 14\,, 22\,, 35$ and $44$ GeV. The uncertainty bands originating from the uncertainty calculations of our {\it residual } charged hadrons FFs also have been presented in these figures. As one can conclude from these figures, the most important effects of the inclusion of {\it residual } charged hadrons FFs are for the case of heavy quark $c$- and $b$-tagged normalized cross sections.

Figure~\ref{fig:data-theory-NLO} shows the comparison of our QCD fit to the fitted SIA data sets at NLO accuracy, while the comparison for the NNLO analysis is shown in Fig.~\ref{fig:data-theory-NNLO}. Notice that in both figures, our results are labeled as ``{\tt Model}''. It should be noted that in these figures the theoretical predictions for cross sections of unidentified light charged hadrons $h^\pm$ have four contributions: the {\it residual} charged hadron contribution determined in our analyses at both NLO and NNLO accuracies and the charged pion, charged kaon and (anti) proton contributions that are determined by the NNPDF collaboration in their NNFF1.0 analysis~\cite{Bertone:2017tyb}. 
In order to present the efficient of the {\it residual} charged hadron FFs in theoretical prediction of unidentified charged hadrons, we compare our results ``{\tt Model}" with the sum of the identified light charge hadrons pion, kaon and (anti) proton from {\tt NNFF1.0} analysis in the following labeled as ``{\tt NNFF1.0}"~\cite{Bertone:2017tyb}.

As one can see from the results presented in these figures, the overall agreement of the SIA experimental data sets in our global QCD analysis of {\it residual } charged hadrons FFs is excellent. All data can be very satisfactorily described by the universal set of {\it residual } charged hadrons FFs determined from this analysis. 
It is clear that in Figs.~\ref{fig:data-theory-NLO} and \ref{fig:data-theory-NNLO}, considering the {\it residual} charged hadron contribution to the unidentified charged hadrons has important role and the theoretical predictions for unidentified charged hadrons by adding the {\it residual} FFs ({\tt Model}) have been improved in comparison with the only pion, kaon and proton FFs ({\tt NNFF1.0}) at both NLO and NNLO accuracies. According to these figures, the most improvements are related to the $c-$ and $b$-tagged normalized cross sections. This finding also is in good agreement with our discussions for the heavy quark FFs and also for the data/theory plots presented in Fig.~\ref{fig:data_theory}.


%
\section{Summary and Conclusions} \label{sec:conclusion}
%

Let us now come to our summary and conclusions. In this paper, we have presented details of a new study of the {\it residual} charged hadrons contributions in unidentified light charged hadrons at NLO and NNLO approximations, which
used experimental information available from single-inclusive unidentified charged hadron production in electron-positron annihilation. The data sets included in our analysis are the {\tt ALEPH}, {\tt OPAL} and {\tt DELPHI} experiments at CERN; the {\tt TPC} and {\tt SLD} experiments at SLAC, and {\tt TASSO} experiment at DESY. 
These data sets were used jointly in both of our NLO and NNLO QCD analyses, allow us to extract the
set of {\it residual} charged hadrons FFs that provides the optimal overall description of the SIA data, along with the estimates of their uncertainties. Since we do not access to  the calculations for the hadronization processes in SIDIS and proton-proton collisions at NNLO, we can not include SIDIS and proton-proton collisions experimental data sets in this analysis.

The unidentified light charged hadron cross sections are total of the identified light charged hadron cross sections, i.e. pion, kaon and (anti) proton, and also the {\it residual} light charged hadron cross sections. Consequently, in order to determine the {\it residual} charged hadron FFs by using the unidentified light charged hadron observables, we need to include the charged pion, charged kaon and (anti) proton FFs. In our analyses we use the light charged hadron FFs of {\tt NNFF1.0} FFs from the NNPDF collaboration.
  
We have presented the techniques, the analyzed data sets, the parameterization and our computational methods of our  {\it residual} charged hadrons FFs analysis. Our technique is formulated in $z$-space using the publicly available {\tt APFEL} code. We have performed uncertainty estimates for our {\it residual} charged hadrons FFs, using the ``Hessian method''. We found that the ``Hessian approach'' yielded consistent results for moderate departures from the best fit, typically for the tolerance of $\Delta \chi^2 = 1$.

With the information from SIA data sets alone, one can only obtain the $q + \bar q$ and gluon FFs. This clearly demonstrates the need for improvements on the FFs from other experiments.
For the future, one can use any observable from hadron productions as well as SIDIS processes. These data sets would give information on the gluon FFs for a wide range of $z$, and also  would provide a clean new probe of the light and heavy {\it residual} charged hadrons FFs.
Our results in this study indicate that there is significant potential for the small {\it residual} charged hadrons contributions in inclusive charged hadrons and considering the small but efficient of  the {\it residual} charged hadrons
improve the agreement between the theoretical predictions and experimental observables. Furthermore, the study presented in this paper has also shown that the {\it residual} contributions become also sizable for the heavy quark FFs as well as the $c$- and $b$-tagged cross sections. To provide further important insights into charged hadrons FFs, it will be straightforward to include all the forthcoming data sets in a certain global QCD analysis.

%
\begin{acknowledgments}
%

Authors are thankful to Rodolfo Sassot and Valerio Bertone for many helpful discussions and comments. We gratefully acknowledge the help and technical assistance provided by Muhammad Goharipour. Authors thank School of Particles and Accelerators, Institute for Research in Fundamental Sciences (IPM) for financial support of this project.
Hamzeh Khanpour also is thankful the University of Science and Technology of Mazandaran for financial support provided for this research.
F. Taghavi-Shahri is grateful Ferdowsi University of Mashhad for financial support for this project. This work is supported
by Ferdowsi University of Mashhad under Grant No.3/46985 (1397/04/19).

\end{acknowledgments}
%


\end{document}